\documentclass[aps,prd,preprint,superscriptaddress]{revtex4-1}
\usepackage{amssymb,amsmath}
\usepackage{graphicx} % Include figure files
\usepackage{dcolumn}  % Align table columns on decimal point
\usepackage{epsfig}   % For PostScript figures
\usepackage{hyperref}
\usepackage{ifpdf} % for handling both LaTeX and PDFLaTeXF
\usepackage{enumitem}

\newcommand{\widebar}[1]{ \overline{#1} }
\newcommand{\tr}{ {\mathbf{tr}\, }}
\newcommand{\Tr}{ {\mathbf{Tr}\, }}

\newcommand{\bra}{\langle}
\newcommand{\ket}{\rangle}
\newsavebox{\accentbox}
\begin{document}

\preprint{}

%\title{An Exact, Finite, Gauge-Invariant, Non-Perturbative Model of QCD Renormalization}
\title{An Exact, Finite, Gauge-Invariant, Non-Perturbative Model of QCD Renormalization}


%\author{H. M. Fried$^{\dag}$, P. H. Tsang$^{\dag}$, Y. Gabellini$^{\ddag}$,  T. Grandou$^{\ddag}$ and Y.-M. Sheu$^{\dag\ddag}$}
\author{H. M. Fried}
\email[]{fried@het.brown.edu}
\affiliation{Physics Department, Brown University, Providence, RI 02912, USA}
\author{P. H. Tsang}
\email[]{Peter\_Tsang@brown.edu}
\affiliation{Physics Department, Brown University, Providence, RI 02912, USA}
\author{Y. Gabellini}
\email[]{yves.gabellini@inln.cnrs.fr}
\affiliation{Universit\'{e} de Nice Sophia-Antipolis, Institut Non Lin$\acute{e}$aire de Nice, UMR 6618 CNRS, 06560 Valbonne, France}
\author{T. Grandou}
\email[]{thierry.grandou@inln.cnrs.fr}
\affiliation{Universit\'{e} de Nice Sophia-Antipolis, Institut Non Lin$\acute{e}$aire de Nice, UMR 6618 CNRS, 06560 Valbonne, France}
\author{Y.-M. Sheu}
\email[]{ymsheu@alumni.brown.edu (corresponding author)}
\affiliation{Physics Department, Brown University, Providence, RI 02912, USA}
\affiliation{Universit\'{e} de Nice Sophia-Antipolis, Institut Non Lin$\acute{e}$aire de Nice, UMR 6618 CNRS, 06560 Valbonne, France}
%\thanks{}
%\affiliation{${}^{\dag}$ {Physics Department, Brown University, Providence, RI 02912, USA} \\ ${}^{\ddag}$ {Universit\'{e} de Nice Sophia-Antipolis, Institut Non Lin$\acute{e}$aire de Nice, UMR 6618 CNRS, 06560 Valbonne, France}}

\date{\today}

\begin{abstract}
A particular choice of renormalization, within the simplifications provided by the non-perturbative property of Effective Locality, leads to a completely finite, renormalized theory of QCD, in which all correlation functions can, in principle, be defined and calculated. In this Model of renormalization, only the Bundle chain-Graphs of the cluster expansion are non-zero. All Bundle graphs connecting to closed quark loops of whatever complexity, and attached to a single quark line, provided no 'self-energy' to that quark line, and hence no effective renormalization. However, the exchange of momentum between one quark line and another, involves only the cluster-expansion's chain graphs, and yields a set of contributions which can be summed and provide a finite color-charge renormalization that can be incorporated into all other QCD processes. An application to High Energy elastic pp scattering is now underway.
\end{abstract}

\pacs{}
%\keywords{}

\maketitle

\section{\label{SEC1}Introduction and Review of References}

A recent, analytic formulation of non-perturbative, gauge-invariant, realistic QCD \cite{FriedGabellini2010,2,3,4,5,6} used elementary functional techniques to sum all gluon exchanges between any pair of quark and/or antiquark lines, including cubic and quartic gluon interactions, yielding a realistic quark-binding potential not approximated by static quarks, as well as (to our knowledge) the first example of nucleon-binding directly from QCD. The result is that individual gluon exchanges disappear from the theory, within their infinite sums each replaced by a \textit{Gluon Bundle} (GB). What remains to be calculated, for any QCD correlation functions, is then the attachment of such GBs to quarks and to all possible combinations of closed-quark-loops (CQLs), as well as the definition of a procedure of quark/hadron renormalization.

It is wise at this point to remember (or at least paraphrase) Schwinger's remark defining renormalization as the change, or more properly a return, from the 'field picture' back to the 'particle picture'. In Abelian QED, for example, where the sum of all radiative corrections defines a dressed photon or lepton propagator, with its $Z_3$ or $Z_2$ factor multiplying the mass-shell pole of that propagator, renormalization is simply defined as the division by, and effective removal of those $Z_{2}$ and $Z_{3}$ factors. But in non-Abelian, gauge-invariant QCD, where the initial gluon fields and propagators disappear from the final analysis, what shall be the physically correct prescription of renormalization? In this paper we formulate one such definition, in which simplicity and finiteness both play a major role. We make no claim to the uniqueness of our definition of renormalization, but only that it is a very simple and obvious way of performing both quark and coupling constant renormalization, in which every step of the program is finite.

For ease of presentation and clarity, we here make two simplifying approximations, which can easily be corrected and extended, as desired. The CQL with but two GBs attached is written in terms of the well-known, un-renormalized, QED lepton loop, with color factors appropriated to QCD appended; in this way, the intrinsic spin dependence of the quark and anti-quark which form the CQL have been included, while there remains to be calculated an extra spin dependence peculiar to QCD. In addition, the transverse arguments $x'$ of each $\left[ f \cdot \chi (x'-v(t')) \right]^{-1}$, representing the connection of each loop to its pair of GBs, will be approximated by the 'averaged' $x_{\mu}$ coordinate appropriate to that loop. As shown in Appendix B of Ref.~\cite{2}, this makes no change in the single-loop amplitude; and a similar statement can be obtained for the chain-loop calculation of this paper.

In order to keep this paper one of finite size, we urge all interested readers to first familiarize themselves with the material of Refs.~\cite{FriedGabellini2010,2,3,4,5,6}. Ref.~\cite{2} explains our method of achieving manifest gauge-invariance, and ends with a non-perturbative description of asymptotic freedom. It also contains what we believe to be the first exact statement of Effective Locality (EL), in which the two endpoint space-time coordinates of every GB are shown to be the same, which property provides tremendous computational simplicity, reducing a basic functional integral (FI) to a set of ordinary integrals~\cite{5}.

Ref.~\cite{2} employs the same functional techniques to display the small absurdities which appear in the amplitudes of all non-perturbative QCD processes -- for example, the exchange of a GB between two quarks -- because use of the standard QCD Lagrangian makes no provision for the experimental fact that asymptotic quarks are always found in bound states; and hence that, in principle, their transverse coordinates can never be measured nor specified precisely. To avoid such difficulty, a 'transverse imprecision' integration over an unknown, transverse probability amplitude,
\begin{equation}
%\displaystyle
\int \mathrm{d}^4 x \bar{\psi}(x) \gamma_{\mu} \lambda_{a} \psi (x) A^a_{\mu}(x)  \rightarrow  \int \mathrm{d}^4 x \int \mathrm{d}^2 x'_{\perp} \mathfrak{a}(x_{\perp} - x'_{\perp}) \bar{\psi}(x') \gamma_{\mu} \lambda_{a} \psi(x')A^a_{\mu}(x)
\end{equation}

\noindent where $ x'_{\mu} = (x_0,x_L,x'_{\perp})$ with $\mathfrak{a}(x'_{\perp}-x_{\perp})$ real and symmetric under the interchange of $x'_{\perp}$ and $x_{\perp}$, is introduced into the quark-gluon part of the Lagrangian, which has the immediate effect of removing all such absurdities. The theory described by this extended Lagrangian is what we have called 'Realistic QCD'.

In Ref.~\cite{2}, a particularly simple choice of the corresponding transverse probability amplitude $\varphi(\vec{b}) = \int \mathrm{d}^{2}\vec{q}\  e^{i \vec{q} \cdot \vec{b}}(\widetilde{\mathfrak{a}}(q))^2$ is inserted into the standard QCD Lagrangian, along with a simple and physically correct method of identifying that effective potential which generates the easily-calculated, non-perturbative eikonal function that would give the desired quark scattering and/or binding; and our result is the potential of form $V(r) \sim \mu(\mu r)^{1+\xi}$, where $\mu$ is the scale parameter for quark binding, on the order of the pion mass, and $\xi$ is a small, real, positive parameter of order $1/10$. By inspection, there are here two parameters essential for quark binding, $\mu$ and $\xi$, rather than just $\mu$ alone.

Ref.~\cite{3} carries the analysis one step further, and provides what we believe to be the first example of Nuclear Physics binding directly from QCD, using our eikonal method to obtain the effective potential which easily binds two nucleons -- two triads of bound quarks -- into a model deuteron. In eikonal and quenched approximations, Refs. \cite{4,5} use an exact Random Matrix calculation to show that the amplitude corresponding to a single GB exchanged between a quark and anti-quark can be written in terms of Meijer G-functions (indeed, a finite sum of finite products of them!), in agreement with general theoretical arguments~\cite{15}; and that both expected SU(3) Casimir invariants contribute to this amplitude, in contrast to lattice-gauge and other model calculations of this amplitude, which contain but one such invariant. The extension to more complicated processes will presumably involve products and integrals over such Meijer G-functions, but this remains to be studied. Finally, Ref.~\cite{6} provides a Brief Review of the first three papers mentioned above.

\section{\label{SEC2}Formulation}

We begin by considering the dressed quark propagator as the simplest example of a QCD correlation function,
\begin{equation}\label{Eq:1}
\mathbf{S}'_c(x-y) = \mathcal{N} \, \int{\mathrm{d}[\chi] \, e^{\frac{i}{4} \, \int{\chi^2}} \cdot \det[f \cdot \chi]^{-\frac{1}{2}} \, \left. e^{\hat{\mathfrak{D}}_A} \, {\mathbf{G}_{\mathrm{c}}(x,y|A) \, e^{\mathbf{L}[A]}}\right|_{A \rightarrow 0} },
\end{equation}

\noindent where $\hat{\mathfrak{D}}_A = \frac{i}{2}\int \frac{\delta}{\delta A} (gf\cdot \chi)^{-1} \frac{\delta}{\delta A}$.  $\mathbf{G}_{\mathrm{c}}[A]$ represents a 'potential theory' quark propagator in the presence of a fictious 'classical' field $A^a_{\mu}(x)$; and $\mathbf{L}[A]$ is the 'CQL functional', representing the sum of a single CQL with all possible (even) numbers of A-fields attached, $\mathbf{L}[A] = \tr{\ln[1 + ig \gamma (\lambda \cdot A) \, \mathbf{G}_c [0] ]}$.  Each $(f\cdot \chi)^{-1}$ factor is associated with the exchange of a specific GB; and the normalization constant $\mathcal{N}$ is the product of the normalization of the Halpern functional integral (FI), divided by the \textit{vacuum expectation value} (VEV) of the QCD S-matrix.

The derivation of Eq.~\eqref{Eq:1}, the relation of $(f\cdot \chi)^{-1}$ to the GB exchanged between any two quark and/or antiquark lines, and the overall gauge-invariant structure is fully and clearly described in Refs.~\cite{FriedGabellini2010,2}. The only point which may require special emphasis is the definition of the measure of FI over the Halpern's field $\chi^a_{\mu\nu}(x)$~\cite{7}, in the conventional sense of breaking up all of space-time into $n$ small 4-volumes of size $\delta^4$ with the understanding that $n \rightarrow \infty$ as $\delta \rightarrow 0$. This FI is NOT to be considered as a sum over all "function space", out of which one may choose one or more convenient examples of $\chi^a_{\mu\nu}(x)$, anti-symmetric in $\mu,\nu$ and carrying a color index $a$.

This distinction becomes crucially important upon realizing that non-perturbative QCD is a theory which contains EL. What this signifies is that -- in contradistinction to its perturbation approximations -- the sum of the gluon exchanges which define each GB take place locally -- $\bra x|GB|y \ket = [g f\cdot \chi(x)]^{-1}\delta^{(4)}(x-y)$ -- so that the relevant Halpern FI can be reduced to an ordinary integration over small $\delta^4$-volume in which the effective interaction occurs. Integrations over all other $\delta^4$-volume elements produce, with their normalization factors, multiple products of $+1$.

\begin{figure}
\ifpdf
\includegraphics[height=35mm]{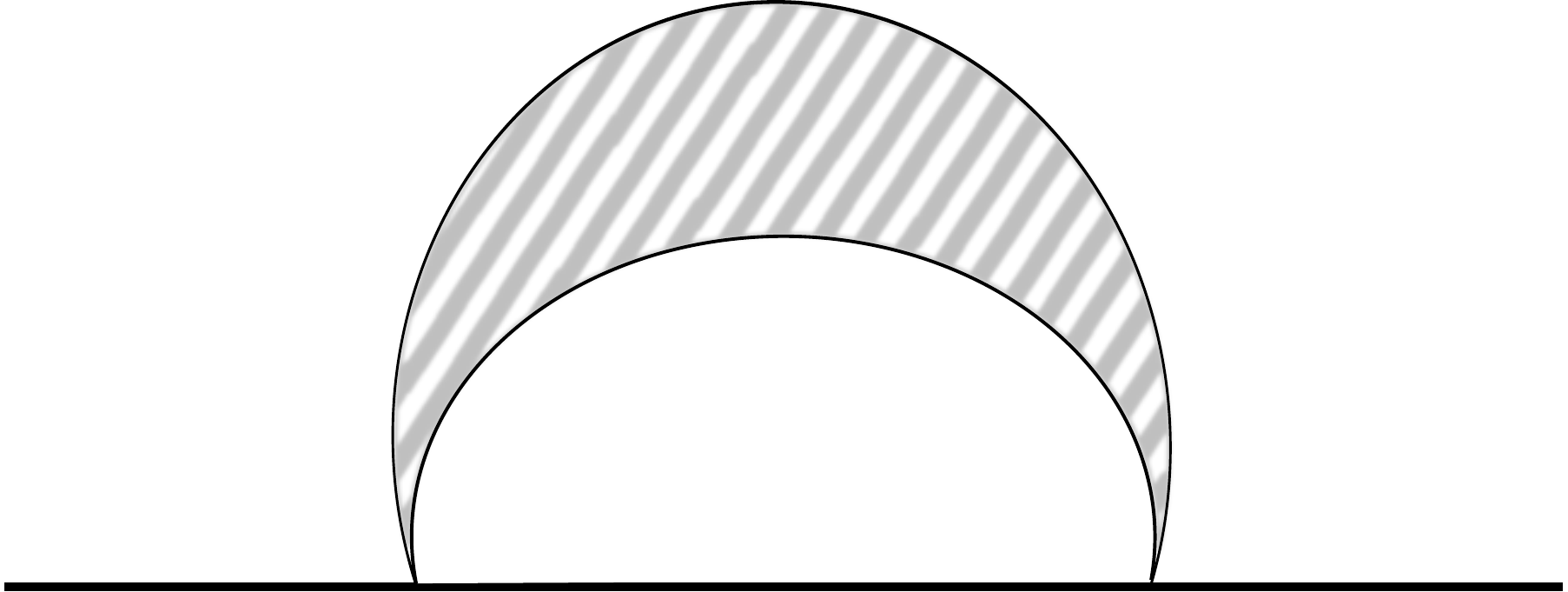}
\else
\includegraphics[height=35mm]{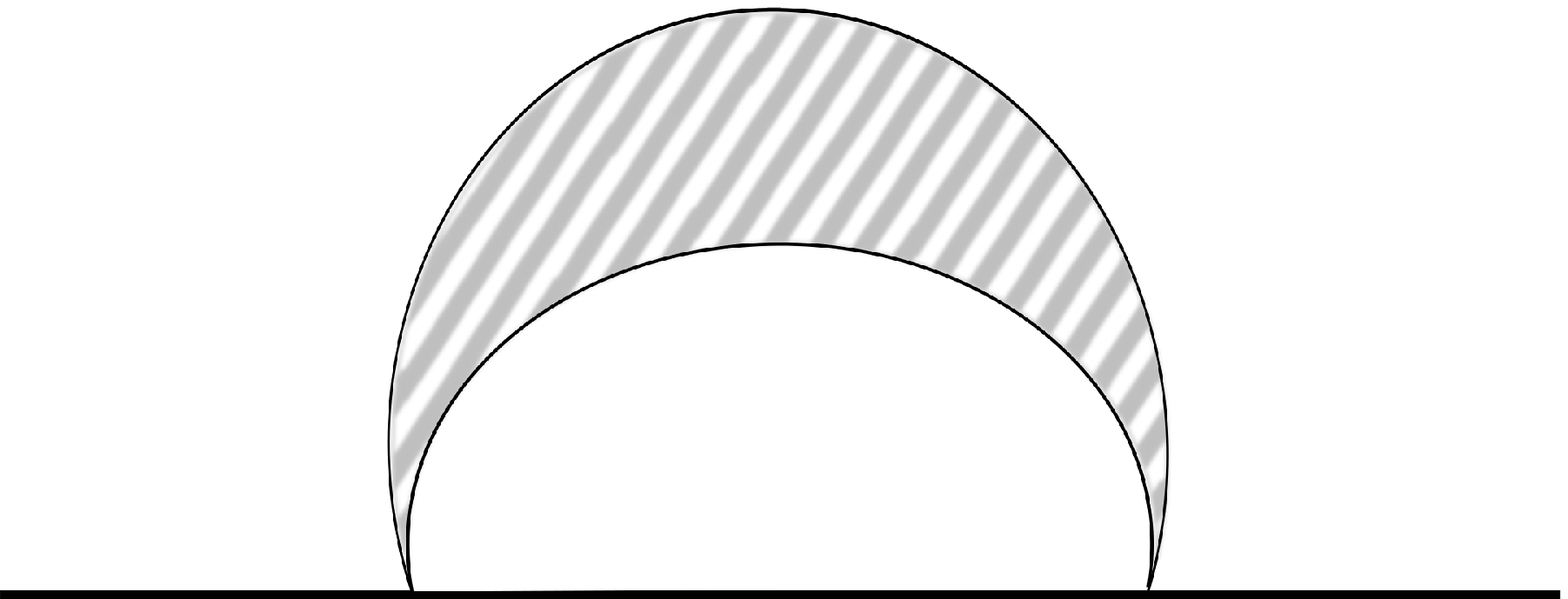}%
\fi
\caption{\label{Fig:1}Gluon Bundle (GB) self-energy (patterned moon shape) across a single quark line (solid line).}
\end{figure}

\begin{figure}
\ifpdf
\includegraphics[height=35mm]{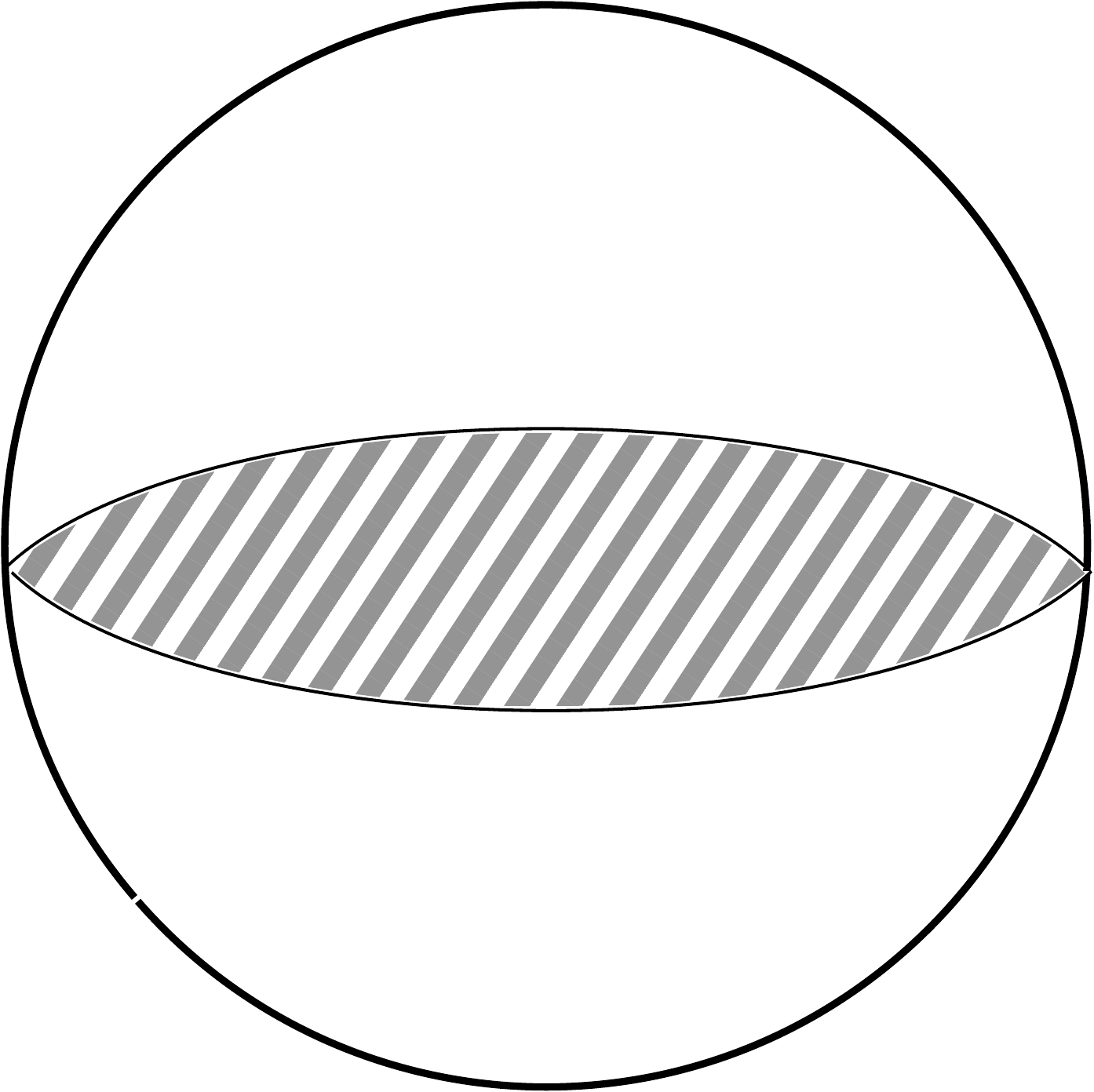}%
\else
\includegraphics[height=35mm]{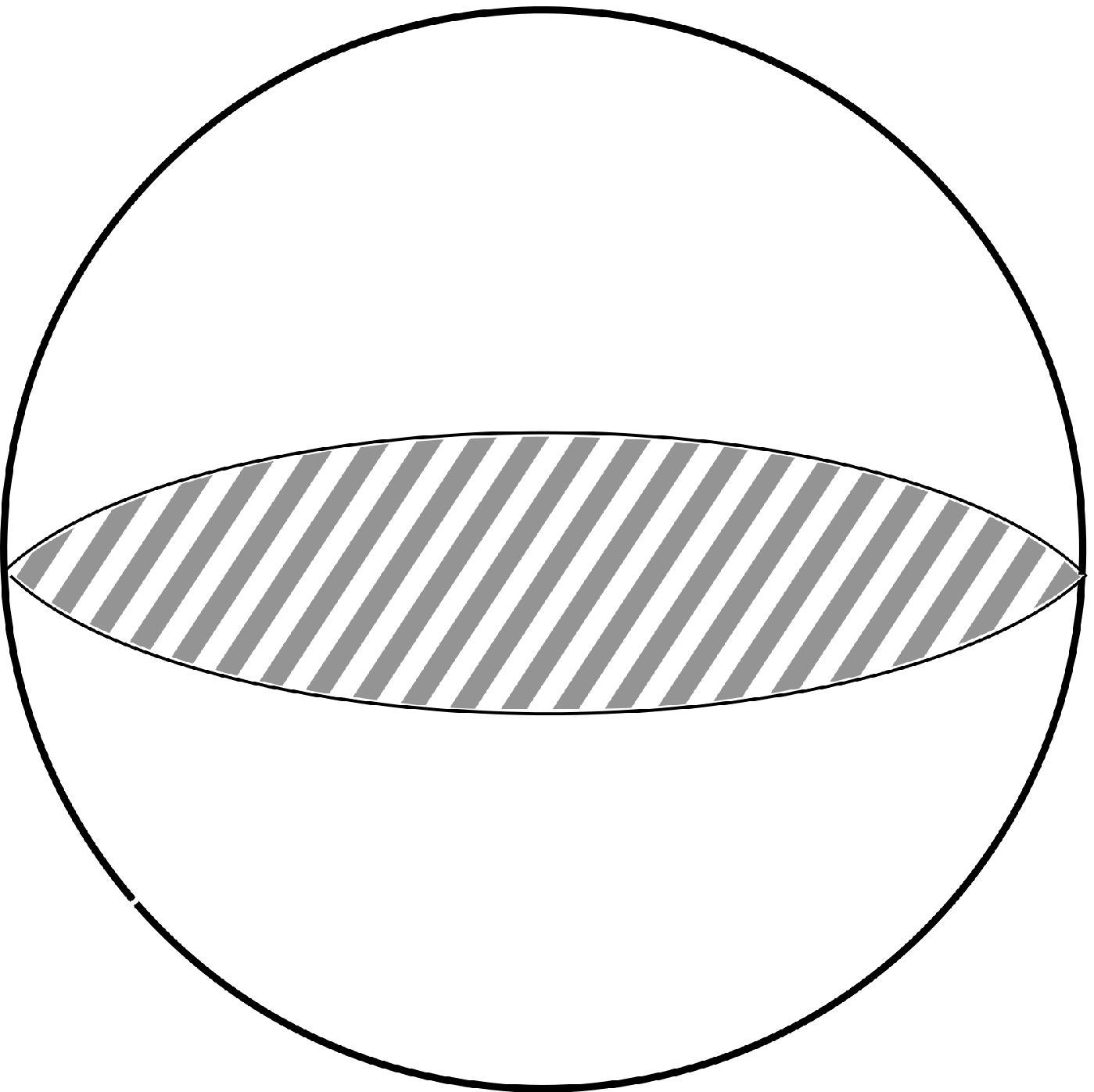}%
\fi
\caption{\label{Fig:2}Closed-quark-loop (solid circular line) with an internal Gluon Bundle (patterned oval shape).}
\end{figure}

In this simplest example of the dressed quark propagator, we shall employ a convenient form of Fradkin's original representations~\cite{8} for $\mathbf{G}_c[A]$ and $\mathbf{L}[A]$, both of which are Gaussian in $A$, and signify that the linkage operations of Eq.~\eqref{Eq:1} can be carried through exactly; for clarity, these representations are reproduced in Appendix~\ref{AppA}. The immediate interest now is the structure of the linkage operator upon the product
\begin{equation}\label{Eq:2}
e^{\hat{\mathfrak{D}}_A} \left( \mathbf{G}_c[A] \, e^{\mathbf{L}[A]} \right) = (e^{\hat{\mathfrak{D}}_A}\, \mathbf{G}_c[A])\, e^{\overleftrightarrow{\mathfrak{D}}_{A}} \, (e^{\hat{\mathfrak{D}}_A} \, e^{\mathbf{L}[A]}),
\end{equation}

\noindent where, with an obvious notation, $\overleftrightarrow{\mathfrak{D}}_A = i \int \frac{\overleftarrow{\delta}}{\delta A}(gf\cdot \chi)^{-1}\frac{\overrightarrow{\delta}}{\delta A}$, and each such $\hat{\mathfrak{D}}_A$ operation has the effect of inserting a GB between the quark propagator $\mathbf{G}_{\mathrm{c}}[A]$ and a CQL functional $\mathbf{L}[A]$, as well as inserting a GB across a quark line, as represented by the "self-energy" graph of Fig.~\ref{Fig:1}, or across the interior of a loop, as in Fig.~\ref{Fig:2}. In ordinary perturbation theory, where an individual gluon would replace the GB above, such graphs are highly divergent.  In realistic QCD, because of Effective Locality (EL), they vanish. For ease of presentation, we have moved the proof of this statement to Appendix~\ref{AppB}, and here continue with the truly relevant part of the formulation, which defines and uses the functional cluster expansion~\cite{9}.

Because of this simplification,
\begin{equation}
(e^{\hat{\mathfrak{D}}_A}\mathbf{G}_{\mathrm{c}}[A])\rightarrow \mathbf{G}_{\mathrm{c}}[A],
\end{equation}

\noindent but the cross-linkage operation
\begin{equation}\label{Eq:3}
(\mathbf{G}_{\mathrm{c}}[A])\, e^{\overleftrightarrow{\mathfrak{D}}_A}\, (e^{{\mathfrak{D}_A}}e^{{\mathbf{L}[A]}})
\end{equation}

\noindent will link the quark propagator with every element of $\mathbf{L}[A]$, in a simple, but important "translational operation" way. If, for simplicity, we momentarily neglect the spin structure of $\mathbf{G}_{\mathrm{c}}[A]$, then all of its $A$-dependence will appear under its defining Fradkin representation integral, in the factor $\exp{\left[-ig \int^s_0{\mathrm{d}s' \, \Omega^{a}(s')\cdot u'_{\mu}(s')\cdot A_{\mu}^{a}(y'-u(s'))} \right]}$, where $u_{\mu}(s')$ is the functional variable whose integration defines the Fradkin representation of $\mathbf{G}_c[A]$, and hence of the dressed quark propagator, while $s$ is the proper-time variable associated with the space-time properties of that propagator.  What this means, in general, is that this $s$-dependence will be inserted into the defining structure of $\mathbf{L}[A]$, which will then exercise a certain measure of control over the subsequently needed $\int_0^{\infty}{\mathrm{d}s \, e^{-ism^2}}$ over the $s$-dependence contributed by $\mathbf{G}_c[A]$ \cite{friedgabelliniannals2012}.

We now turn to the functional cluster expansion, defined combinatorially and picturally in Ref. \cite{9}, which takes the form
\begin{equation}\label{Eq:3a}
e^{{\hat{\mathfrak{D}}_A}} \, e^{{\mathbf{L}[A]}} = e^{\sum_{\ell=1}^{\infty} \frac{Q_\ell}{\ell!} }
\end{equation}

\noindent where $Q_{\ell} = e^{\hat{\mathfrak{D}}_A}(\mathbf{L}[A])^{\ell} |_{\mathrm{conn}}$, and where the subscript 'conn' means that only multiple loops attached to each other by at least one GB are retained. For example,
\begin{equation}
Q_1[A] = e^{\hat{\mathfrak{D}}_A}\mathbf{L}[A] \equiv \widebar{\mathbf{L}}[A]
\end{equation}

\noindent and
\begin{equation}
Q_2 = \widebar{\mathbf{L}}[A] \, (e^{\overleftrightarrow{\mathfrak{D}}} - 1) \, \widebar{\mathbf{L}}[A]
\end{equation}

\noindent etc.  But, as noted in Appendix~\ref{AppB}, every GB inserted across the same loop will always vanish, and therefore $\widebar{\mathbf{L}}[A] \rightarrow \mathbf{L}[A]$. The multiplicative linkages of all the $Q_{\ell}$ then correspond to all possible GB insertions between different loops, with their 'self-energies' missing.

%
% Figures are floating in RevTeX/LaTeX
%
\begin{figure}
\ifpdf
\includegraphics[height=35mm]{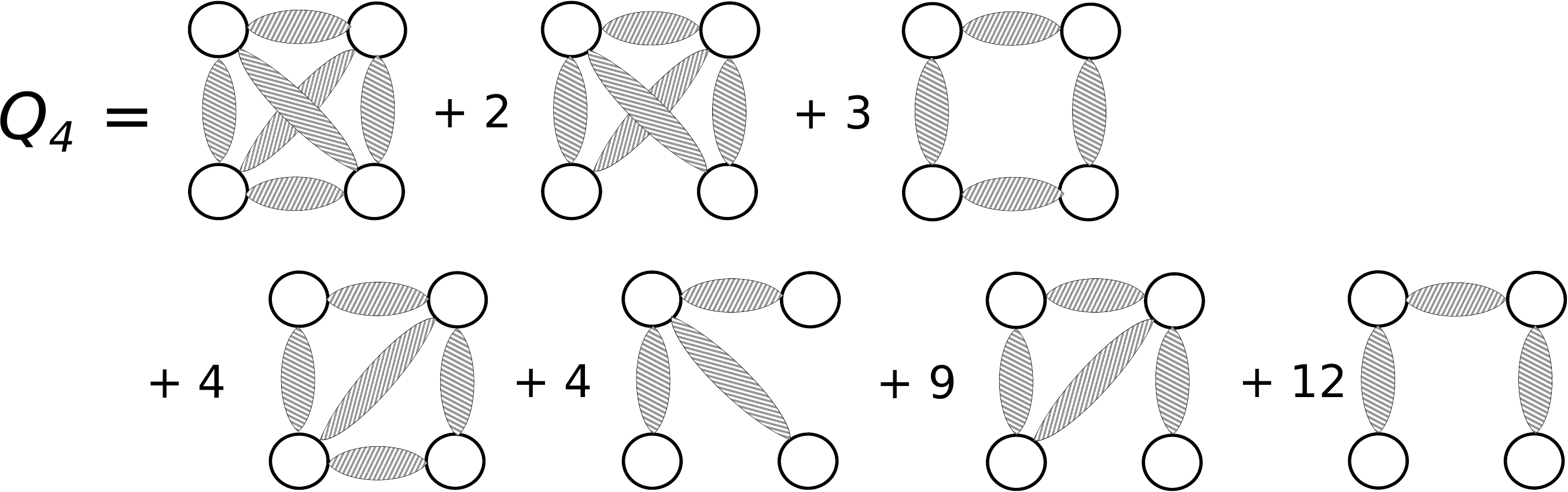}%
\else
\includegraphics[height=35mm]{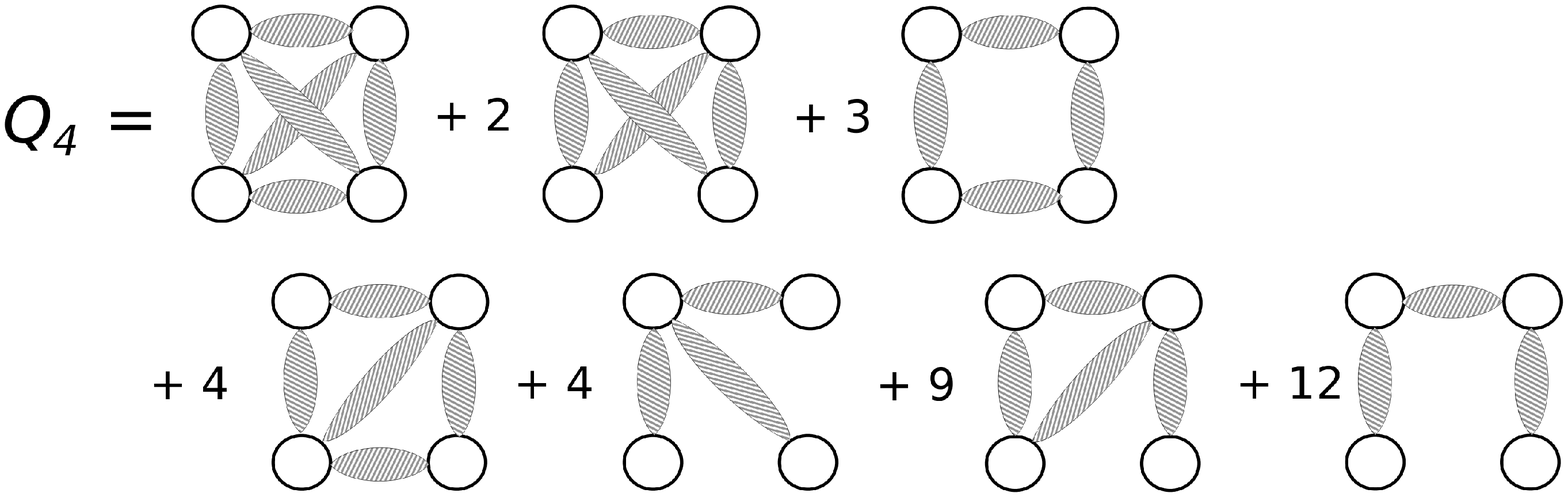}%
\fi
\caption{\label{Fig:3}$Q_4$ term in the cluster expansion (solid circles as loops and patterned ovals as GBs).}
\end{figure}

As an example, consider a pictorial representation of $Q_4$, taken from Ref.~\cite{9}, and reproduced here as Fig.~\ref{Fig:3}, where the patterned ovals represent all possible numbers of GB linkages between the loops, and the integers on the left-side of each graph represent the statistical weight of that arrangement of closed loops. The cross-linkages between $\exp\{{Q_{\ell}}\}$ and $\mathbf{G}_c[A]$ will require two additional GBs linking the quark line with each of the diagrams of $Q_{\ell}$, in all possible ways.

One immediate simplification is provided by the fact that the Fradkin representations for any loop will be non-zero for an even number of GB attachments to that loop. But there are then still a huge number of possible linkages of $Q_{\ell}$ to $\mathbf{G}_{\mathrm{c}}[A]$; and the reduction to just a few such linkages, which can be easily summed, is the goal of the next sections.

\section{\label{SEC3}QUARK RENORMALIZATION}

It will be most efficient at this point to temporally restrict attention to the class of 'chain graphs', such as the last RHS loop combination of Fig.~\ref{Fig:3}. It has a statistical weight of $12 = \frac{4!}{2}$, and this numerator factor of $4!$ serves to cancel the $\frac{1}{4!}$ in the sum of Eq.~\eqref{Eq:3a}, leaving behind a net multiplicative factor of $\frac{1}{2}$. And such cancellation holds for every chain graph, for every value of $\ell$.

\begin{figure}
\ifpdf
\includegraphics[height=35mm]{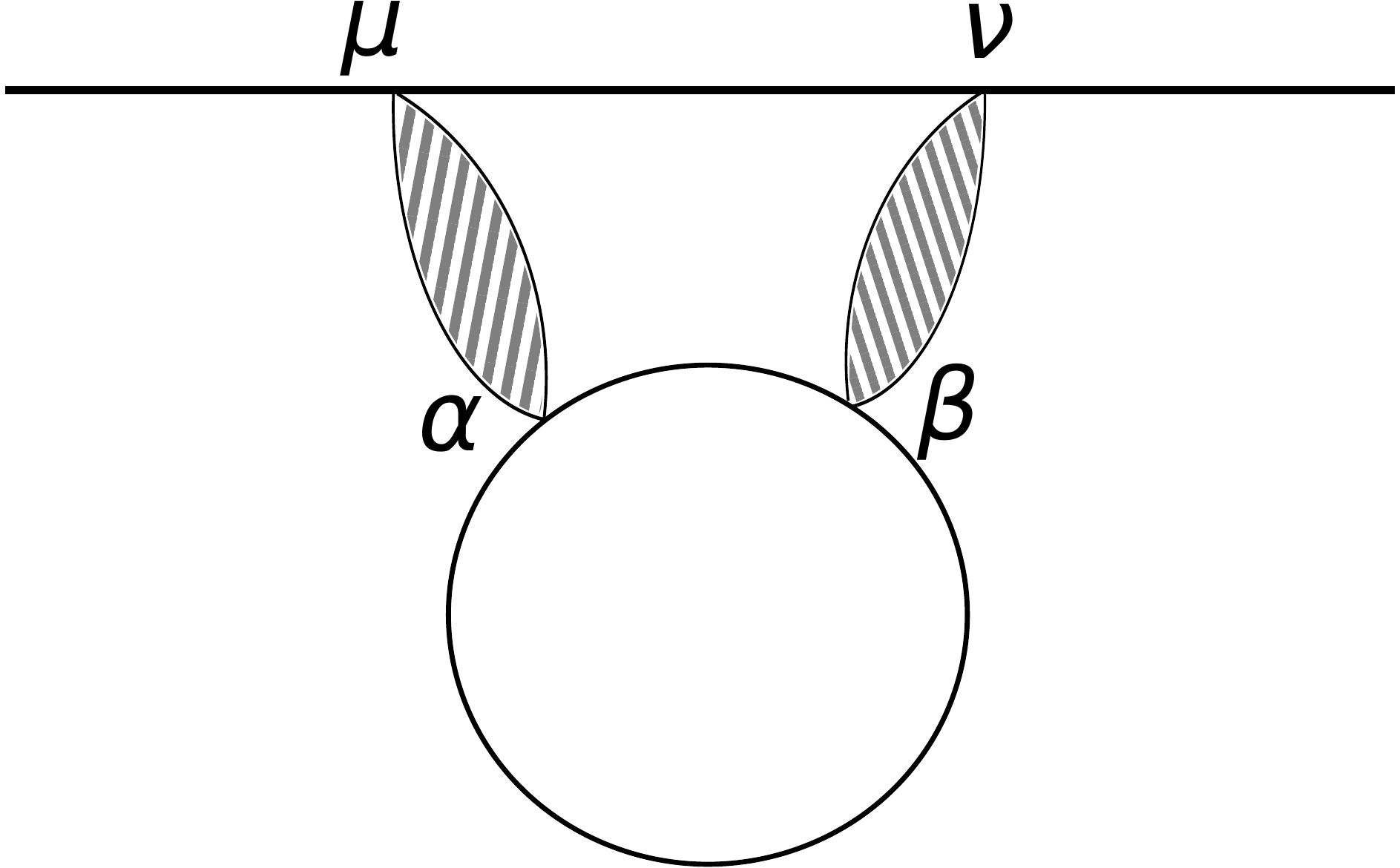}
\else
\includegraphics[height=35mm]{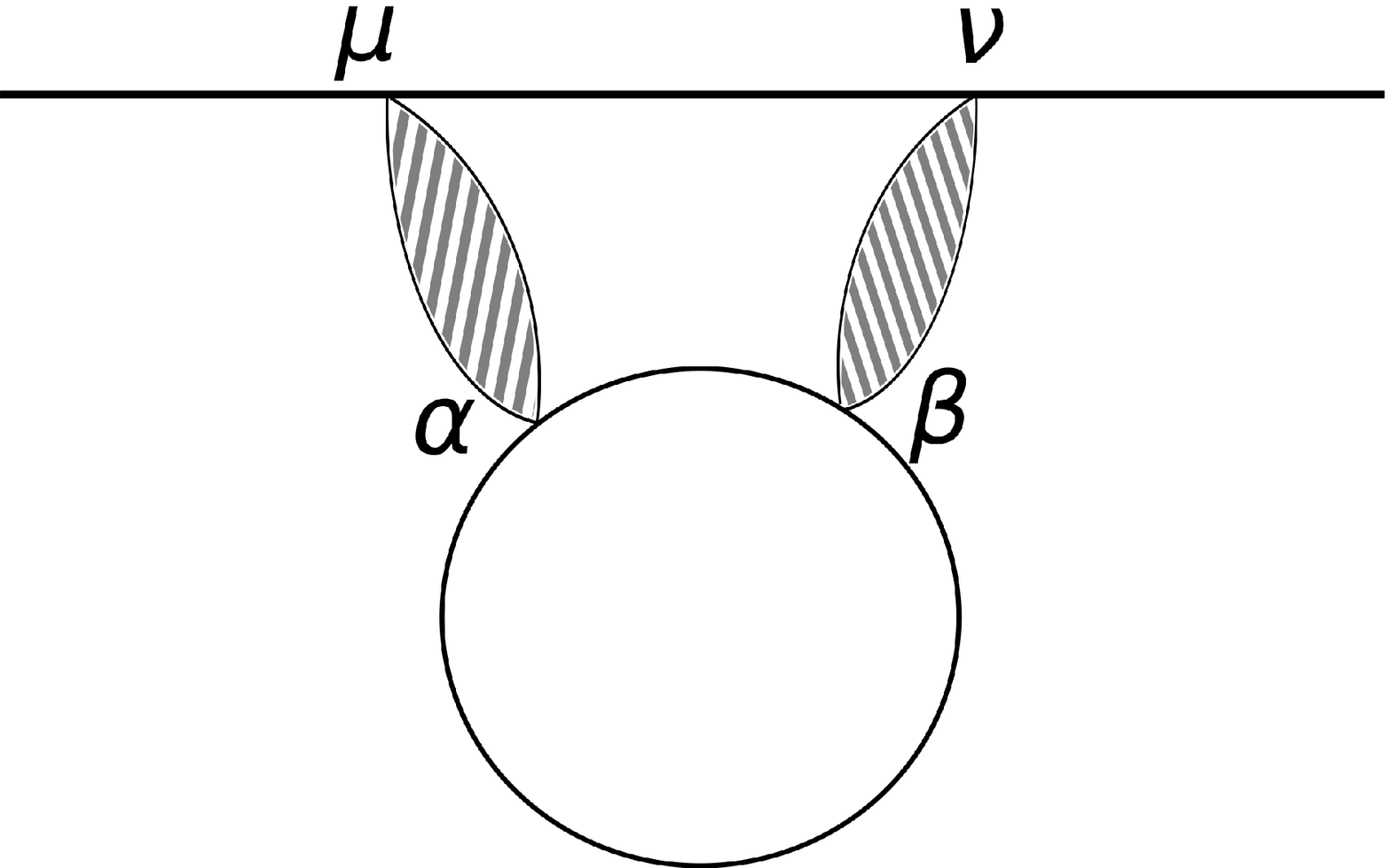}%
\fi
\caption{\label{Fig:4}Simplest radiative corrections corresponding to the chain graph attached to a single quark line.}
\end{figure}

The simplest radiative corrections corresponding to the chain graphs attached to a single quark line are pictured in Fig.~\ref{Fig:4}. where there are two sorts of terms which enter the FI over the closed loop, one due to the spin of virtual quarks, and the other due to the non-spin $\int_0^s{\mathrm{d}s' \, u'(s') \, \Omega(s') \, \int_0^t{\mathrm{d}t' \, v'(t') \, \widehat{\Omega}(t')}}$ terms which multiply the attached GBs. For simplicity, we here outline the calculation of the second contribution, and then, in words, simply state the more obvious result of the spin dependence.

Each of the two GBs of the loop of Fig.~\ref{Fig:4} is proportional to a 4-dimensional delta-function of their end-point variables,
\begin{equation}\label{Eq:4}
v'_{\alpha}(t_1) \cdot \delta^{(4)}(y' - u(s_1) - x' + v(t_1)) \cdot \delta^{(4)}(x'' - v(t_2) - y' + u(s_2)) \cdot v'_{\beta}(t_2),
\end{equation}

\noindent where $v(t')$ is the space-time coordinate of the loop whose FI defines its Fradkin representation, and, as always, $z'_{\mu} = (z_0, z_L, z'_{\perp})$. At every intersection of a GB with a quark line there will appear a transverse integration over the relevant probability amplitudes, in this case $\int{\mathrm{d}^{2} x'_{\perp} \, \mathfrak{a}(x_{\perp}-x'_{\perp})} \cdot \int{\mathrm{d}^2 x''_{\perp}\, \mathfrak{a}(x_{\perp}-x''_{\perp})} \cdot \int{\mathrm{d}^2y' \, \mathfrak{a}(y_{\perp}-y'_{\perp})} \cdot \int{\mathrm{d}^{2}y'' \, \mathfrak{a}(y_{\perp} - y''_{\perp})}$ and the generalization to higher numbers of loops forming the chain graphs is immediate.

As a first step in the calculation, it will be useful to consider the product of the time-like and longitudinal $\delta$-functions of Eq.~\eqref{Eq:4}, $\delta^{(0,L)}(y-u(s_1)-x+v(t_1))\cdot \delta^{(0,L)}(x-v(t_2)-y+u(s_2))$, for upon integration over the corresponding loop coordinates $\int{\mathrm{d}x_0 \, \int{\mathrm{d}x_L }}$ (required by the Fradkin representation), one obtains the product $\delta^{(0)}\cdot \delta^{(L)}$, or more compactly,
\begin{equation}\label{Eq:5}
\delta^{(0,L)}(u(s_1)-u(s_2)-v(t_1)+v(t_2)),
\end{equation}

\noindent and we evaluate both the $\delta^{(0)}$ and $\delta^{(L)}$ of Eq.~\ref{Eq:5} by assuming that there are a set of points, $t_{\ell}$ for the first and $t_m$ for the second, about which $v_0(t_1)$ and $v_L(t_2)$ can be expanded -- $v_0(t_1)$ about $t_{\ell}$, and $v_L(t_2)$ about $t_m$ -- at which point the arguments of each $\delta$-function vanishes. The result is, for the product of both functions,
\begin{equation}\label{Eq:6}
\delta^{(0,L)} \rightarrow \sum_{\ell} \frac{\delta(t_1-t_{\ell})}{|v'_{0}(t_{\ell})|} \cdot \sum_{m} \frac{\delta(t_2-t_m)}{|v'_{L}(t_m)|}\bigg|_{\substack{v_0(t_{\ell})-v_0(t_m)=u_0(s_1)-u_0(s_2) \\ v_L(t_{\ell})-v_L(t_m)=u_L(s_1)-u_L(s_2)}}.
\end{equation}

%
% missing
%
Since $v_0$ and $v_L$, and $u_0$ and $u_L$, are completely arbitrary continuous functions of their variables, each possessing a first derivative -- while the set of $C^{1}[(0,s) \rightarrow \mathbb{R}]$-functions is of measure zero in the Wiener-space relevant to the $u$, $v$-functions -- the probability of finding points $t_\ell$ and $t_m$ to fit the subsidiary conditions of Eq.~(\ref{Eq:6}) is arbitrary small.  The only values of $t_{\ell, m}$ which can satisfy these conditions are $t_{\ell, m} = $ $0$ or $t$, where we \underline{know} that the pair of restrictions involving the difference of $u_{0, L}(s_1)$ and $u_{0, L}(s_2)$ are satisfied by construction.

But now that $t_1$ and $t_2$ are either $0$ or $t$, upon averaging all the transverse fluctuations, and integrating over the $x_{\perp}$-dependence of the loop, the pair of transverse $\delta$-functions of Eq.~(\ref{Eq:4}) lose all their $v_{\perp}(t_1)$ and $v_{\perp}(t_2)$-dependence, since $v_{\perp}(0) = v_{\perp}(t) = 0$. Immediately, the FI of the loop over its $v$-dependence will then vanish,
\begin{equation}
\int{\mathrm{d}[v] \, v'_{\alpha}(t_1) \, v'_{\beta}(t_2) \, e^{\frac{i}{2} \int{v \cdot (2h)^{-1} \cdot v} } \, \delta^{(4)}(v(t)) } = 0,
\end{equation}

\noindent and the contribution of this Bundle Graph to the quark's '\emph{dressing}' is zero. The same result appears for the purely spin contributions to the Fradkin representation of the loop, because these are given in terms of gradients of y-dependence, which dependence vanishes from the product of the $\delta$-functions of Eq.~(\ref{Eq:4}), after the transverse fluctuations and the $x_{\perp}$ loop variables are integrated out.

\begin{figure}
\ifpdf
\includegraphics[height=35mm]{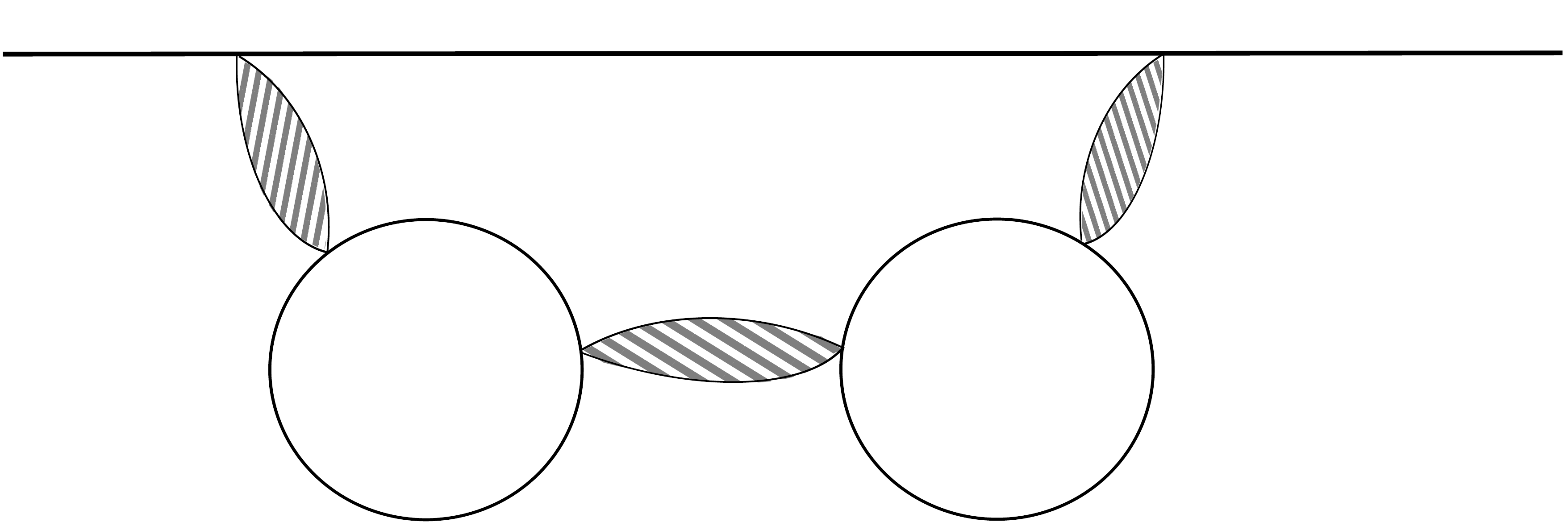}%
\else
\includegraphics[height=35mm]{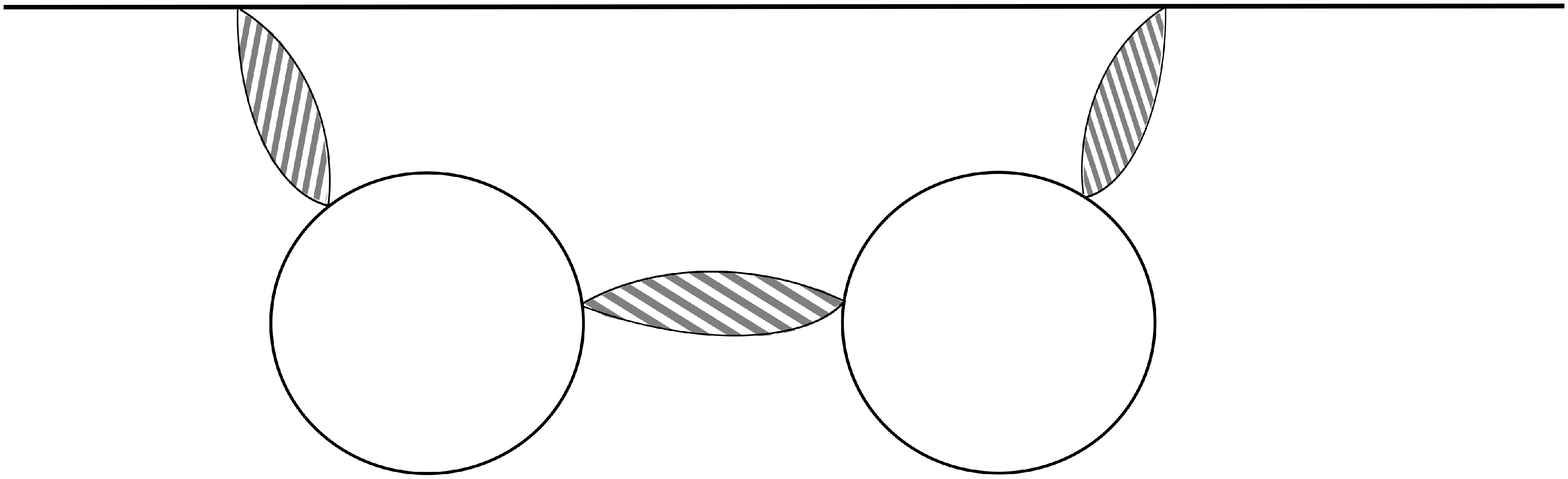}%
\fi
\caption{\label{Fig:5}Higher-loop Chain Bundle Graphs across a single quark line.}
\end{figure}

Higher-loop Chain Bundle Graphs, such as those pictured in Fig.~\ref{Fig:5}, will also vanish.  The analysis is simplest if one applies the above arguments to the 'end-loops' -- those carrying a GB attached to the quark line, as in Fig.~\ref{Fig:5} -- which will always vanish, independently of the number of loops in the chain. In this way, one sees that the Chain Bundle Graphs do not contribute to the quark 'self-energy'; they do not contribute to the 'dressing' of the quark propagator. However, the Bundle Graphs $\underline{will}$ contribute to the processes involving momentum transfer between two quarks and/or antiquarks; and these subprocesses will provide the basis for a finite color-charge renormalization.

\section{\label{SEC4}GLUON BUNDLE RENORMALIZATION}
%\subsection{}
%\subsubsection{}

We now give a definition of GB renormalization appropriate to the current situation in which individual gluon exchange has already been summed, and all gluons have effectively disappeared from the problem. This definition is most useful because it removes \underline{all} of the many, non-chain Bundle Graphs in the cluster expansion of Section~\ref{SEC2}.

Remember that the measure of the Halpern integral is composed of a normalized product of $n$ very small 4-volume elements, $\delta^4$, which span the entire 4-volume, with the understanding of the subsequent limits, $n \rightarrow \infty$ and $\delta \rightarrow 0$. The Gaussian weighting of the FI can be written as
\begin{equation}\label{Eq:7}
\prod_{n}{\mathcal{N}_n \, \int{\mathrm{d}[\chi] \, e^{\frac{i}{4} \, \delta^4  \, \chi^2_n} \, \det{[f \cdot \chi_n ]^{-\frac{1}{2}}} \, \mathcal{F}[(f\cdot \chi_n)^{-1}]}},
\end{equation}

\noindent where $\chi_n = \chi(x_n)$, the subscript $n$ labels the small space-time volume in which the variable $\chi^a_{\mu\nu}(x_n)$ is defined; and that $x_n$ variable is to be integrated over all possible real values, from $-\infty$ to $+\infty$, independently of all other $x_{m\neq n}$ values. In Eq.~(\ref{Eq:7}), $\mathcal{F}$ contains the exponentiated dependence characteristic of the exchange of a GB, and $\mathcal{N}_n$ is the normalization of that $n$-th integral, such that its value is unity when $g \rightarrow 0$.

Let us now change to a new, dimensionless variable $\bar{\chi}$, defined by $\delta^2 \chi = \bar{\chi}$, so that $\chi^{-1} = \delta^2 \, \bar{\chi}^{-1}$. With this trivial change, the normalizations $\mathcal{N}_n$ are now independent of $\delta$, while the exponential interaction term now carries a multiplicative factor of $\delta^2$, since $(f\cdot \chi)^{-1} \rightarrow \delta^2(f\cdot\bar{\chi})^{-1}$. In pictorial terms, every GB now carries a $\delta^2$ factor, which may be imagined as a single $\delta$ factor appearing at each end of the GB.

We now define "GB renormalization" in the following way, effectively paraphrasing Schwinger's comment that renormalization is what must be done in returning from the 'field picture' to the 'particle picture', where in this case the particle is the quark. (It may always be asymptotically bound -- as distinct from asymptotically free -- but it is still the physical 'particle' in QCD.)

When a GB connects to a quark, one which is or will eventually be bound, asymptotically, into a hadron, the $\delta$ at that end of the GB is to be replaced by a real, finite, non-zero $\delta_q$. But the $\delta$ at the other end of that GB, connecting a virtual quark loop -- which is not a physical particle -- is to be maintained as a factor which is subsequently going to vanish. The renormalization of that infinitesimal $\delta$ connected to the quark line can be viewed as similar to the removal of the wave-function renormalization $Z$ factors multiplying the "free particle" part of a dressed propagator in conventional perturbation expansions; both the $Z$, whose inverse contains one or more UV divergences, and the $\delta$, are redefined to obtain the renormalized forms of each. And, as in the conventional theory, when one "divides" by that $Z$ factor to obtain the renormalized propagator, one is "dividing by zero", to effectively replace $Z$ by 1.

However, $\delta_q$ has a dimension of length or time, and appropriate care must be taken in assigning it a numerical value. Since we expect the most significant contribution to any such high-energy scattering process to appear when the CM quark space-time indices are either 4 or 3, corresponding to energy or to a related longitudinal momentum, we might well permit Quantum Mechanics (QM) to make the choice for us, replacing $\delta_q$ by a factor proportional to $\frac{1}{E} \approx \frac{1}{p_3}$.  This introduces a reasonable energy dependence into the amplitude, which will act in such a way -- as the Center of Mass (CM) energy increases -- to decrease the QM interference between separate chains linking the scattering quarks. Of course, this assumption must be verified by comparison with extensive pp scattering data, at energies ranging from GeV to TeV; and this will be explored in a separate analysis.

Now consider one loop of a Chain Bundle Graph whose Fradkin functional integral $\int{\mathrm{d}[v]}$ is evaluated. Each end of the two GBs which connects that loop contribute a factor of $\delta$, so that a net factor of $\delta^2 \rightarrow 0$ multplies that loop.  Were there no net momentum transfer away from a single quark line, as in the above closed-quark-loop (CQL) analysis of the radiative corrections to a single quark, then that Fradkin functional integral vanishes.  But if there is momentum transfer $q$ passing through that loop, that integral has a well-defined, non-zero dependence on $q$, as well as a logarithmically-divergent UV factor which we shall call $\ell$. Since $\delta^2$ is to vanish, and $\ell$ is to diverge, in their respective limits, and since they appear multiplied together, we define the combination $\kappa = \delta^2 \, \ell$ as a real, finite, positive constant, whose value is to be determined subsequently.

This definition is not unique; but it has the great advantage that Chain Bundle Graphs which transfer momentum produce a finite contribution to their sum; and most importantly, all of the other loops of the cluster expansion vanish: a loop connected to four GBs is proportional to $\delta^4 \, \ell$, while a loop connected to more than four GBs has no log divergence, and contributions of both groups vanish. By the above definition of GB renormalization, only the chain Bundle Graphs survive; and the essentially geometric sum of all of their contributions is able to generate what might appropriately be termed a 'finite color-charge renormalization'. This Model definition is quite appropriate to our nucleon-nucleon binding potential to form a model deuteron, as in Ref.~\cite{3}, where the $\delta^2 \, \ell$ product is replaced by the finite $\kappa$, combined with the coupling, and determined by the ground-state binding energy.

\begin{figure}
\ifpdf
\includegraphics[height=35mm]{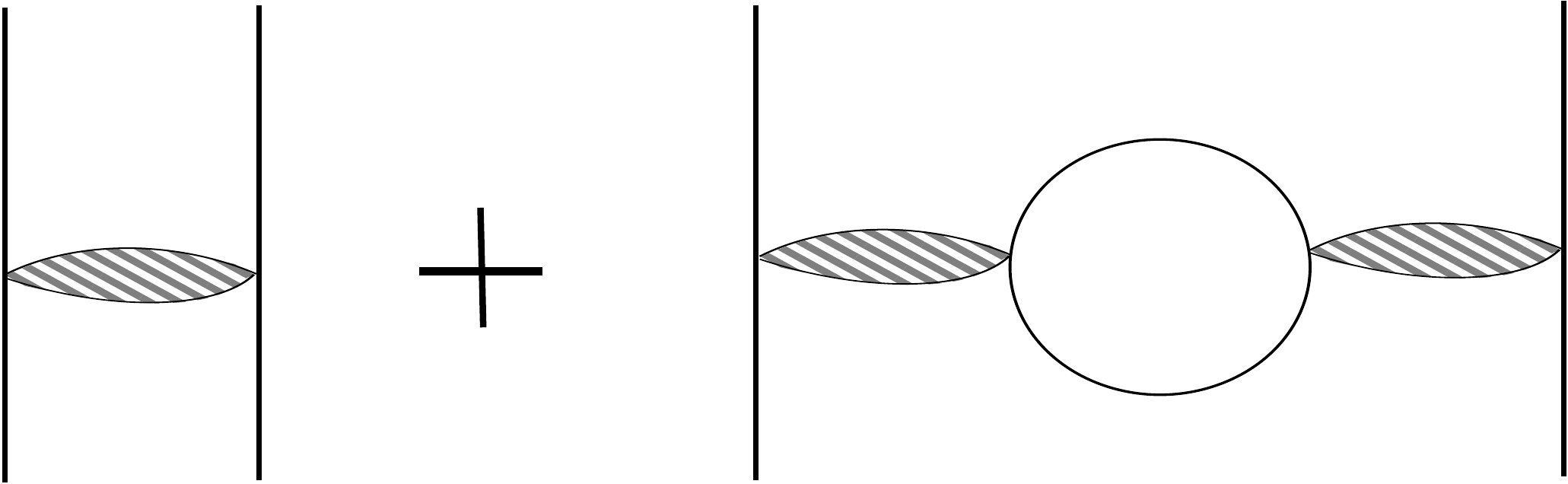}%
\else
\includegraphics[height=35mm]{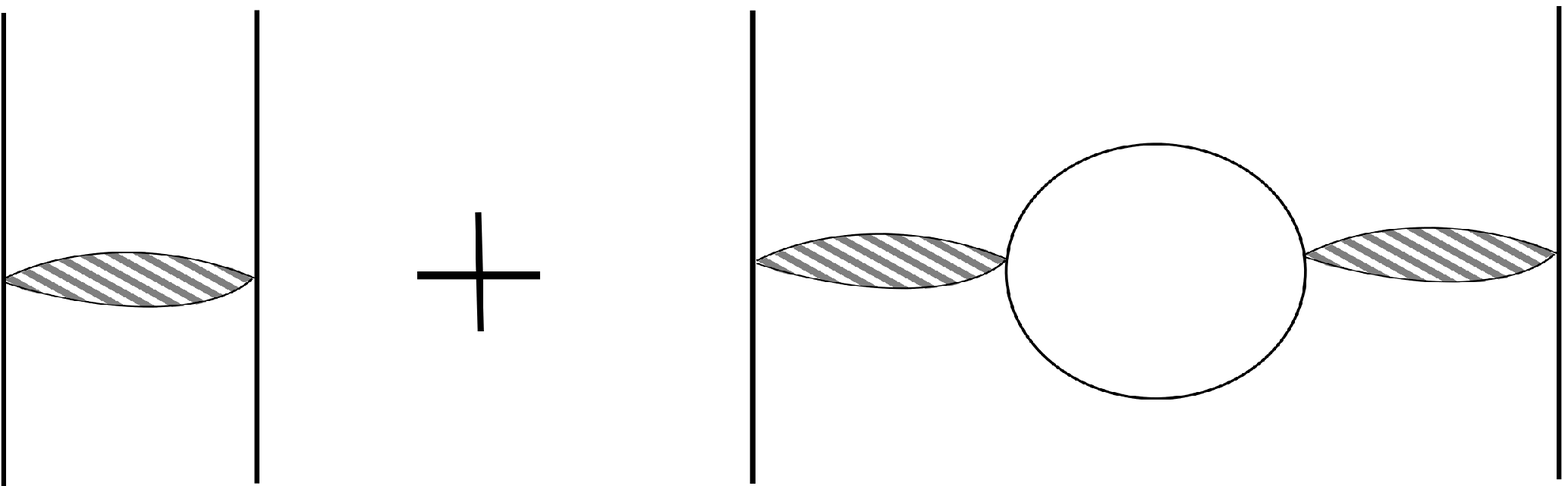}%
\fi
\caption{\label{Fig:7}A single GB plus a single closed loop exchanged between two quarks contributions to nucleon-nucleon scattering.}
\end{figure}

\begin{figure}
\ifpdf
\includegraphics[height=95mm]{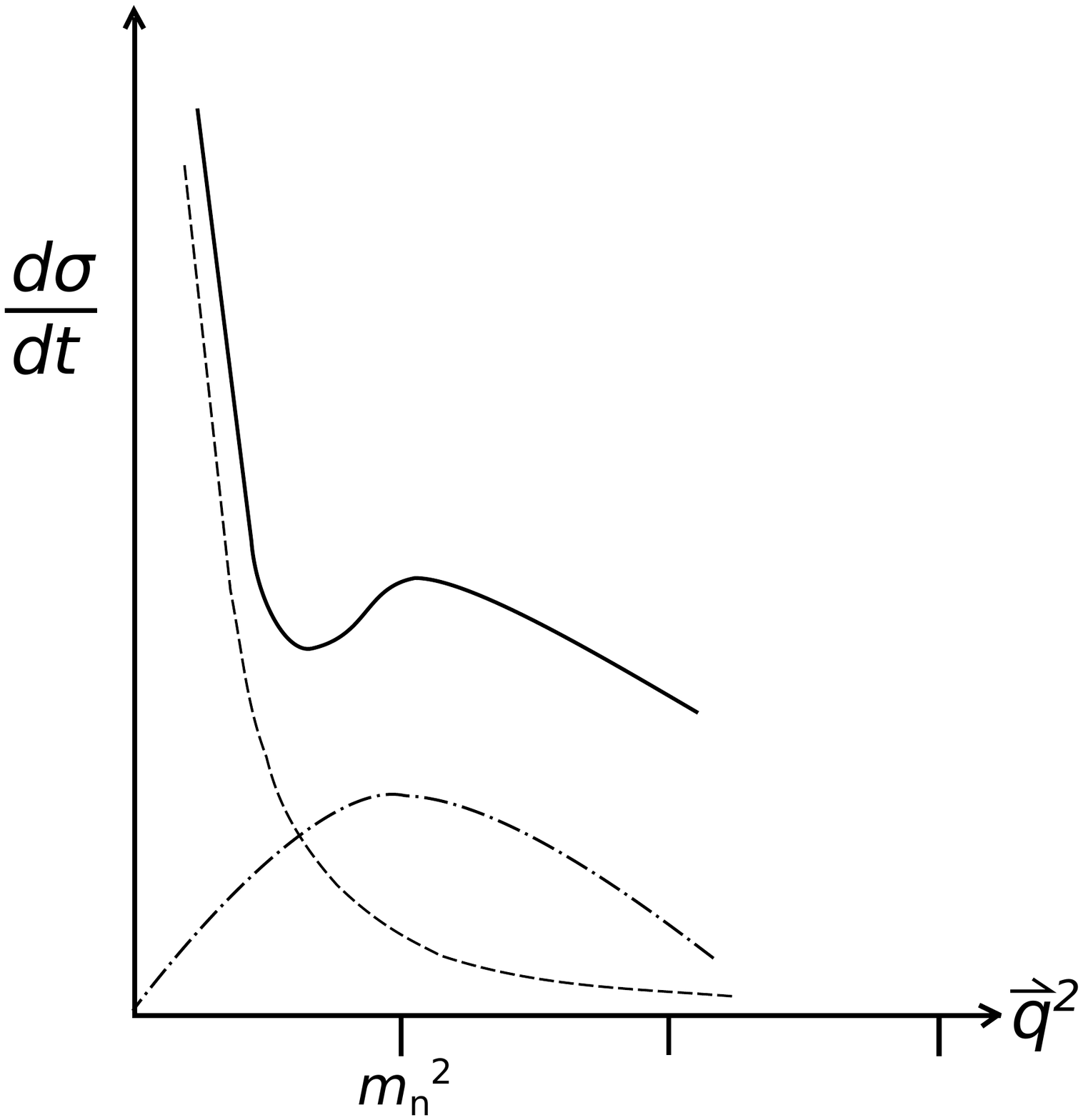}%
\else
\includegraphics[height=95mm]{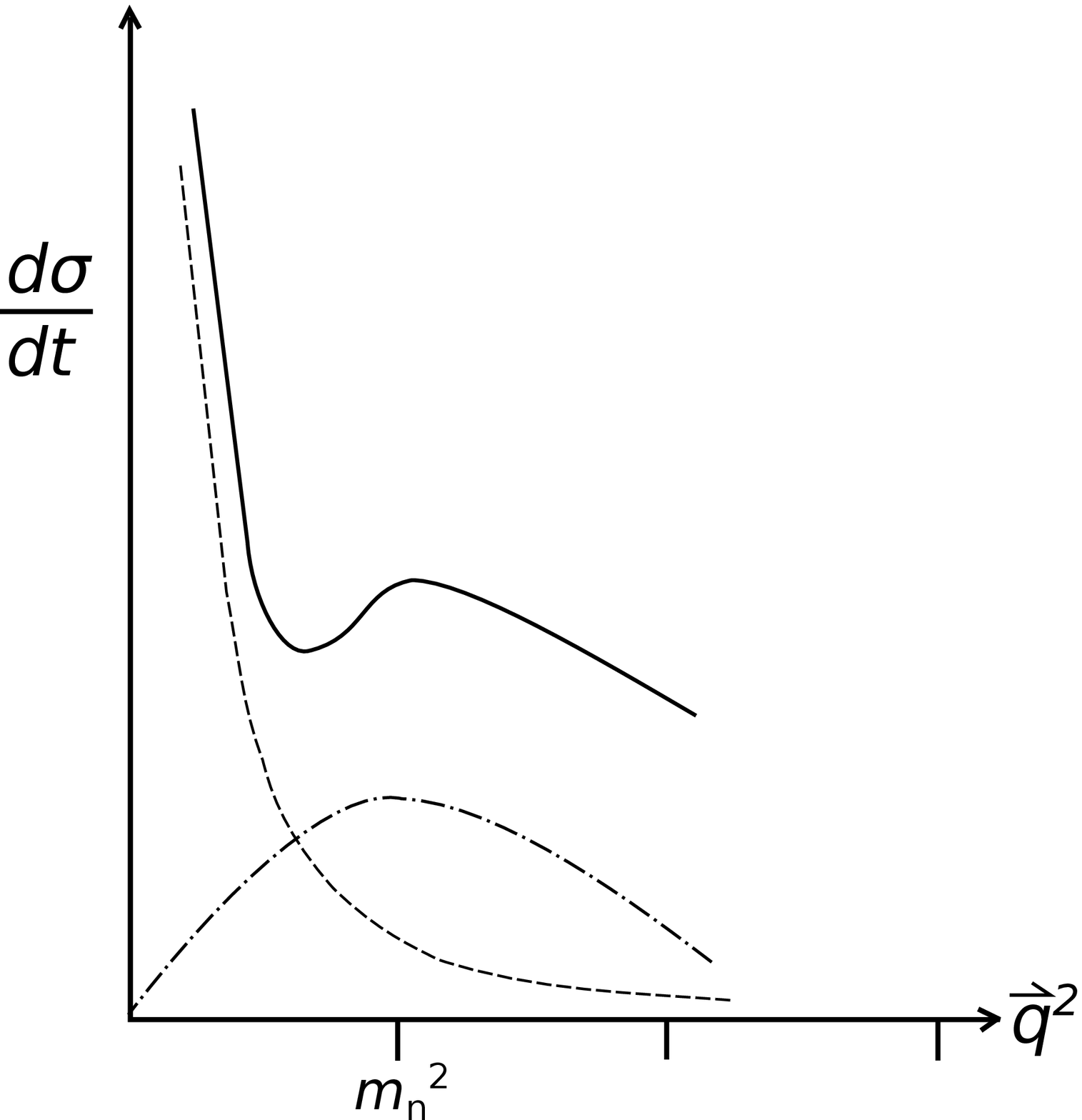}%
\fi
\caption{\label{Fig:8}The differential cross-section of two scattering nucleons.}
\end{figure}

Because such renormalized couplings will show a strong fall-off with increasing $\vec{q}^{\, 2}_{\perp}$, this property suggests an immediate application to the extensive experimental data of pp differential cross-sections at high energies~\cite{totem2011,totem2013}. Using the very approximate replacement of $(f\cdot \chi)$ factors by magnitudes $R$, neglecting all angular correlations between $\chi$-projections in color space, one can easily evaluate an amplitude corresponding to the sum of a single GB plus a single closed loop exchanged between two quarks, as in Fig.~\ref{Fig:7}.  One finds the qualitative result of Fig.~\ref{Fig:8}  for the $\frac{d\sigma}{dt}$ of two scattering nucleons, suppressing all dependence on quark binding which produced those nucleons.

The horizontal axis of Fig.~\ref{Fig:8} is given in units of nucleon mass $m$, while the scale of the vertical axis is arbitrary. The dashed line descending rapidly for small $\vec{q}^{\, 2}_{\perp}$ is due to the single GB exchange, while the dotted line rising from the origin represents one-loop exchange. The $|\mbox{absolute\ value}|^2$ of their sum is given by the solid line, and is of interest because it clearly shows the "diffraction dip" in the region of $0.5 m_n$, followed by the hump and subsequent descent at a higher $\vec{q}^{\, 2}_{\perp}$ value. When more loops are included, the dip should very slightly exceed $0.5 m_n$, while the decrease following the hump will be less rapid. Both features are exhibited by pp scattering data for $\vec{q}^{\, 2}_{\perp}$ values well past the Coulomb interference region. Work is currently underway to refine these calculations, and form a precise representation of the shape of this curve, in particular, well past the hump; but it is reassuring that the approximation evaluation used here shows a strong, qualitative resemblance to the experimental data.

\section{\label{SEC5}SIMPLIFYING THE CHAIN-LOOP CONTRIBUTIONS}
%\subsection{}
%\subsubsection{}

But before that stage can be realized, it is important to point out one property that the careful reader will observe, in this qualitative presentation, where the only quark spin-dependence retained comes from the quark forming the loop -- which has the same form as that of a QED closed fermion loop, with the addition of appropriate color factors -- instead of the more correct result that follows from including the complete spin dependence of that loop, which can be inferred from the exact Fradkin representations of Appendix~\ref{AppA}. The complete and relevant Halpern sub-integral over that portion of the "interior" loops, those lying between the "end-point" or "exterior" loops that connect to the scattering quarks, will also have a role to play in this analysis.

The point to be made here is that the detailed loop and Halpern integral computations must generate a result in which momentum transfer across each loop, and across the sum of all loops, is such that the momentum transfer leaving one quark is received by the other quark, an obvious necessity, but one which is hidden by the details of the computations. What shall be done here is to simplify matters, adopting a simplified form of the loop result, in which this property is guaranteed. Proper orders of magnitude of the $q$-dependence are maintained in this simplification, which guarantees momentum-transfer conservation.

Specifically, suppose that a momentum transfer $q$, moving left-to-right, enters an 'interior' loop which bears the overall, space-time matrix indices of $\alpha$ and $\beta$, corresponding to a momentum $q$ entering the loops as $q_{\alpha}^{(\mathrm{I})}$ on its left-hand-side, and exiting as $q^{(\mathrm{I\!I})}_{\beta}$ on its right-hand-side. These are transverse momenta of two components, \emph{e.g.}, $q^{(\mathrm{I})}_{\alpha}$ and $q^{(\mathrm{I\!I})}_{\beta}$ with $\alpha, \beta = 1, 2$; and if this momentum is going to be transferred across the loop, then the result of the exact Fradkin FI of the loop, together with the exact Halpern sub-integral over the very small space-time interval in which that integral is defined, must combine to produce the effective statement that $q_1^{(\mathrm{I})} = q_1^{(\mathrm{I\!I})}$, and that $q^{(\mathrm{I})}_2 = q^{(\mathrm{I\!I})}_2$. In other words, the indices $\alpha$ and $\beta$ are not arbitrary, but, in effect, must be the same.

The Fradkin FI of a corresponding QED loop has the form~\cite{friedgabelliniannals2012}
\begin{equation}\label{Eq:8}
(q_{\alpha} \, q_{\beta} - \delta_{\alpha\beta}\, q^2) \, \Pi(q^2),
\end{equation}

\noindent where
\begin{equation}\label{Eq:8b}
\Pi(q^2) = \int_0^{\infty}{\frac{\mathrm{d}t}{t} \, e^{-it m^2} \, \frac{e^2}{2\pi^2} \, \int_{0}^{1}{\mathrm{d}y \, y(1-y) \, e^{-it q^2 y(1-y)} }},
\end{equation}

\noindent and contains an obvious logarithmic UV divergence, coming from the behavior of the integrand near its lower limit. This differs from the proper QCD loop integral which contains color-factors, with indices $a'$ and $b'$, associated with locations connected to GBs on each side of the loop, $\int_{0}^{t}{\mathrm{d}t_1 \, v'_{\alpha}(t_1) \, \Omega_{a'}(t_1)} \cdot \int_{0}^{t}{\mathrm{d}t_2  \, v'_{\beta}(t_2) \, \Omega_{b'}(t_2)}$, where the $\alpha$, $\beta$, $a'$, $b'$ indices are joined to neighboring GBs. There is also another QCD quark-spin contribution, which is tied to the Halpern integral in a moderately complicated way. It should also be mentioned that in the product of any two neighboring loops, the factors of $q_{\alpha} q_{\beta} q_{\gamma} q_{\delta}$, appearing in the product of two of the neighboring brackets of Eq.~(\ref{Eq:8}), will give no contribution because the symmetric combination $q_{\beta} q_{\gamma}$ will multiply the antisymmetric -- in space-time and color indices -- factor $(f\cdot\chi)^{-1}|_{\beta\gamma}$ between those loops.

The simplification noted above and now made is simply to retain only the $-\delta_{\alpha\beta}q^2$ factor of Eq.~(\ref{Eq:8}), multiplied by a parameter $\lambda$ which is to represent the result of the Fradkin and Halpern integrations, for as noted above their evaluations must produce such a $\delta_{\alpha\beta}q^2$ factor. The parameter $\lambda$ will multiply the constant $\kappa$, and their product will enter into the definition of the renormalized charge. Since we shall extract only the divergent part of each such loop -- which provides a finite contribution in the limit as the width of the Halpern sub-integral vanishes -- it turns out that the color indices $a'$, $b'$ across each loop are also going to be the same. The fact that both transverse and both color indices of interior loops are the same will generate a significant simplification in the final result.

\begin{figure}
\ifpdf
\includegraphics[height=35mm]{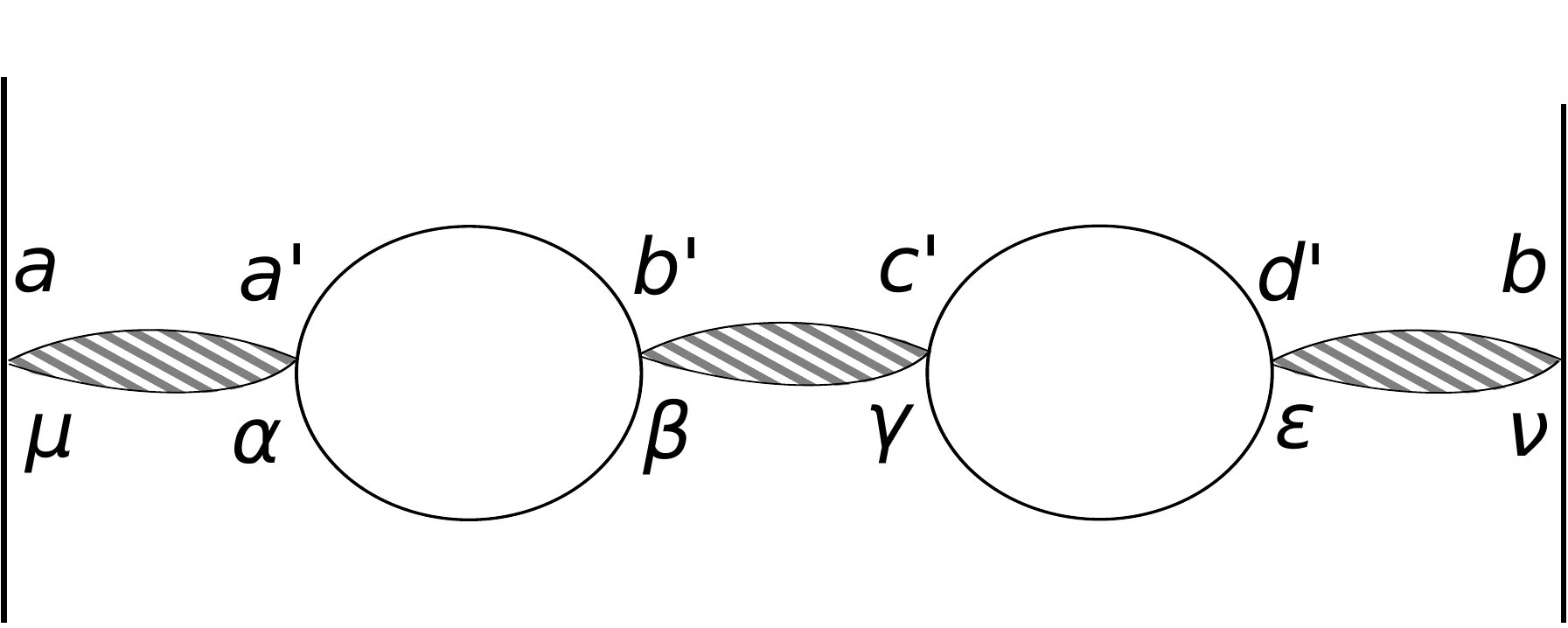}%
\else
\includegraphics[height=35mm]{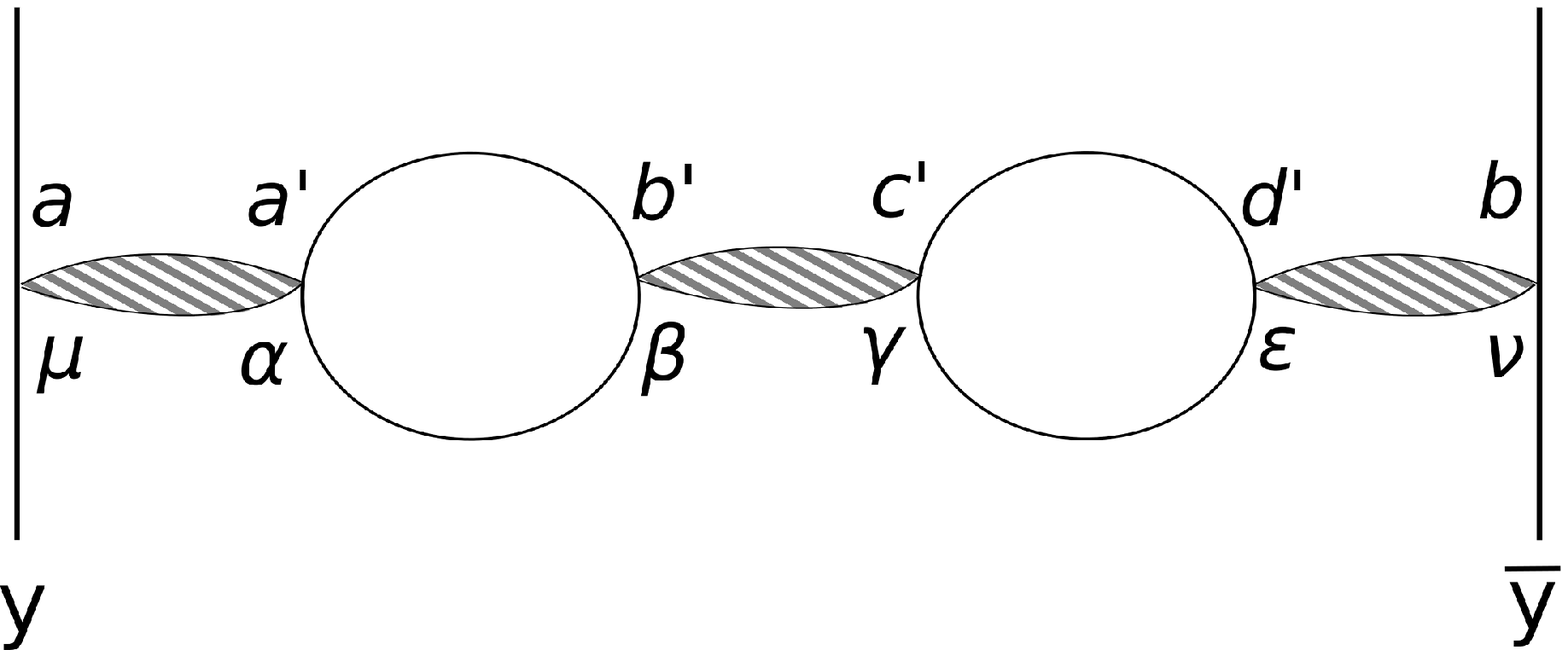}%
\fi
\caption{\label{Fig:6} Quark-Quark interaction with chain graph of two loops.}
\end{figure}

Perhaps the simplest approach is to first consider the two-loop amplitude of Fig.~\ref{Fig:6}, writing only the factors needed for this evaluation,
and to then state a sequence of operations that can be easily performed, along with their results. The amplitude for this process is proportional to the factors
\begin{eqnarray}\label{Eq:9}
&& \int_{0}^{s}{\mathrm{d}s_{1} \, u'_{\mu}(s_1)\, \Omega_{a}(s_1)} \, \int_{0}^{\bar{s}}{\mathrm{d}\bar{s}_{1} \, \bar{u}'_{\nu}(\bar{s}_{1}) \, \widebar{\Omega}_b(\bar{s}_{1})} \cdot \int_{0}^{t}{\mathrm{d}t_1 \, v'_{\alpha}(t_1) \, \widehat{\Omega}_{a'}(t_1)}
\\ \nonumber &\cdot&
\int_{0}^{t}{\mathrm{d}t_2 \, v'_{\beta}(t_2) \, \widehat{\Omega}_{b'}(t_2)} \cdot \int_{0}^{\bar{t}}{\mathrm{d}\bar{t}_1 \, \bar{v}'_{\gamma} \, \breve{\Omega}_{c'}(\bar{t}_1)} \cdot \int_{0}^{\bar{t}}{\mathrm{d}\bar{t}_2 \, v'_{\epsilon}(\bar{t}_2) \, \breve{\Omega}_{d'}(\bar{t_2})}
\\ \nonumber &\cdot&
\int{\mathrm{d}^{2} y'_{\perp}\, \mathfrak{a}(y_{\perp} - y'_{\perp})} \cdot
\int{\mathrm{d}^2 \bar{y}'_{\perp} \, \mathfrak{a}(\bar{y}_{\perp} - \bar{y}'_{\perp})} \cdot
\int{\mathrm{d}^2 x'_{\perp} \, \mathfrak{a}(x_{\perp} - x'_{\perp})}
\\ \nonumber &\cdot&
\int{\mathrm{d}^2 x''_{\perp} \, \mathfrak{a}(x_{\perp}-x''_{\perp})} \cdot
\int{\mathrm{d}^2 \bar{x}'_{\perp} \, \mathfrak{a}(\bar{x}_{\perp} - \bar{x}'_{\perp}) } \cdot
\int{\mathrm{d}^2 \bar{x}''_{\perp} \, \mathfrak{a}(\bar{x}_{\perp}-\bar{x}''_{\perp}) }
\\ \nonumber &\cdot&
\delta^{(4)}(y' - u(s_1) - x' + v(t_1)) \cdot
\delta^{(4)}(x'' - v(t_2) - \bar{x}' + \bar{v}(\bar{t}_1) ) \cdot
\delta^{(4)}(\bar{x}'' - \bar{v}(\bar{t}_2) - \bar{y}' + \bar{u}(\bar{s}_1)  )
\\ \nonumber &\cdot&
[f \cdot \chi(y' - u(s_1))]^{-1}|^{aa'}_{\mu\alpha} \cdot
[f \cdot \chi(\bar{x}' - \bar{v}(\bar{t}_1))]^{-1} |^{b'c'}_{\beta\gamma} \cdot
[f \cdot \chi(\bar{y}' - \bar{u}(\bar{s}_1))]^{-1} |^{d'b}_{\epsilon\nu},
\end{eqnarray}

\noindent where the $\mathfrak{a}(z_{\perp}-z_{\perp}')$ represent the probability amplitudes of each quark to be found at a perpendicular distance $z'_{\perp}$ close to its 'average', or 'Abelian' value $z_{\perp}$;
the square of the 2-D Fourier transform of this quantity, $\tilde{\varphi}(q)= [\tilde{\mathfrak{a}}(q)]^2$, represents the probability of an individual GB event delivering a momentum transfer $q$. The three, 4-dimensional delta-functions of Eq.~(\ref{Eq:9}) are the statement of \emph{Effective Locality} for each of the three GBs, and the primes on their arguments correspond to $z'_{\mu}=(z_0,z_L;z'_{\perp})$, where the subscripts $0$ and $L$ signify time-like and longitudinal components, respectively. The three $(f\cdot \chi)^{-1}$ correspond to the three GBs of this problem, while the $x$ and $\bar{x}$ coordinates are the space-time coordinates of each loop (which must be integrated over); and $y$ and $\bar{y}$ represent the coordinates of each quark, with $u(s')$ and $u(s')$ their Fradkin functional variables. The forms written in this Section are appropriate to the simplest situation of a single GB chain exchanged between the scattering quarks; the most general formulation of multiple GB chains exchanged between the quarks is noted in Section~\ref{SEC7}.

Our simplified and justifiable prescriptions are as follows:
\begin{enumerate}[label=(\alph*)]
\item Suppress the primes in the arguments of each $(f\cdot \chi)^{-1}$; the justification for this step is given in Appendix B of Ref.~\cite{2}.
\item Assume that $[f \cdot \chi(\bar{x} - \bar{v}(\bar{t}_1))]^{-1}$ is labeled only by its transverse arguments, an assumption made for convenience, which is consistent with the final results of this exercise.
\item Write an integral representation for each of the time-like and longitudinal $\delta$-functions of Eq.~(\ref{Eq:9}), thereby introducing the Fourier variables $q_0,q_L,p_0,p_L,k_0,k_L$. Exactly as shown in Section 3 of Ref.~\cite{3}, assume the two quarks of Fig.~\ref{Fig:6} are scattering at high energy, and adopt a simple, Eikonal Model description of that amplitude; this %very good%
    approximation removes the need for an integration over the Fradkin $u$- and $\bar{u}$-dependence. It then follows that all of the Fourier variables $q_0, q_L, p_0, p_L, k_0, k_L$ vanish, so that only transverse $q_{\perp},p_{\perp},k_{\perp}$ dependence is relevant.
\item Write Fourier representations for the remaining three transverse delta-functions of Eq.~(\ref{Eq:9}), and calculate the integrals $\int \mathrm{d}^2 y'_{\perp} \cdot \int \mathrm{d}^2 \bar{y}'_{\perp} \cdot \int \mathrm{d}^2 x'_{\perp} \cdot \int \mathrm{d}^2 x''_{\perp} \cdot \int \mathrm{d}^2 \bar{x}'_{\perp} \cdot \int \mathrm{d}^2 \bar{x}''_{\perp}$ to obtain factors of $\tilde{\varphi}(q)\cdot \tilde{\varphi}(p) \cdot \tilde{\varphi}(k)$ where all previous $z'_{\perp}$ are effectively replaced by $z_{\perp}$.
\item Calculate $\int \mathrm{d}^2 x_{\perp} \cdot \int \mathrm{d}^2 \bar{x}_{\perp}$  and find that $p_{\perp} = k_{\perp}= q_{\perp}$, so that there is but one transverse integral, $\int{\mathrm{d}^{2}q_{\perp} \, } \equiv \int{\mathrm{d}^{2}q \, }$, which remains.
\end{enumerate}

One final question remains: How is one to understand and represent $[f \cdot \chi(\bar{x}-\bar{v}(t_1))]^{-1}$?  The three transverse $\delta$-functions multiplying the last line of Eq.~(\ref{Eq:9}) can be used to re-write this term as
\begin{eqnarray}\label{Eq:10}
&& \left[ f \cdot \chi\left(\frac{1}{2}[x-v(t_2)+\bar{x}-\bar{v}(\bar{t}_{1})]\right) \right]^{-1}
\\ \nonumber &\Rightarrow&
\left[ f \cdot \chi\left(\frac{1}{2}[y-u(s_1)+v(t_1)-v(t_2)-\bar{v}(\bar{t}_{1}) +\bar{v}(\bar{t}_{2})+\bar{y} -\bar{u}(\bar{s_1}) ] \right) \right]^{-1},
\end{eqnarray}

\noindent and, as explained in Ref.~\cite{3}, in the Center of Mass (CM) of the scattering quarks, with the zero of time chosen as that time when both quarks' longitudinal coordinates are zero, the Eikonal Model effectively replaces $y-u(s_1)$ by $y_{\perp}$, and $\bar{y}-\bar{u}(\bar{s}_1)$ by $\bar{y}_{\perp}$. This replaces Eq.~(\ref{Eq:10}) by
\begin{equation}\label{Eq:11}
\left[ f \cdot \chi\left(\frac{1}{2}[y_{\perp} + \bar{y}_{\perp} + \Delta v - \Delta \bar{v}] \right) \right]^{-1},
\end{equation}

\noindent where $\Delta v = v(t_1) - v(t_2)$, and $\Delta \bar{v} = \bar{v}(\bar{t}_1) - \bar{v}(\bar{t}_2)$. And because the CM value of the transverse vectors $y_{\perp} + \bar{y}_{\perp} = 0$, Eq.~(\ref{Eq:11}) reduces to the simpler form
\begin{equation}\label{Eq:12}
\left[ f \cdot \chi\left(\frac{1}{2}[\Delta v - \Delta \bar{v}] \right) \right]^{-1}.
\end{equation}

In contrast, the remaining transverse integral over $\mathrm{d}^2q_{\perp}$ has as its integrand the factors \begin{equation}
e^{i q \cdot [y_{\perp}-\bar{y}_{\perp}+\Delta v + \Delta\bar{v}]},
\end{equation}

\noindent where $y_{\perp}-\bar{y}_{\perp} = \vec{b}$, the impact parameter. The $(f \cdot \chi)^{-1}$ of Eq.~(\ref{Eq:12}) must now be included as part of Fradkin's $v$ and $\bar{v}$-integrals. For this, we write a Fourier representation of Eq.~(\ref{Eq:12}) as
\begin{equation}
\int{\frac{\mathrm{d}^{2}K}{(2\pi)^2} \, \widetilde{\mathcal{F}}(K) \, e^{i \frac{K}{2} \, [\Delta v - \Delta \bar{v}]}},
\end{equation}

\noindent and immediately note that the UV divergent part of the Fradkin integrals over both loops, $\int{\mathrm{d}[v]} \cdot \int{\mathrm{d}[\bar{v}]}$, is proportional to the product
\begin{equation}\label{Eq:13}
\left[ -\lambda \, \delta_{\alpha\beta} \, (q + \frac{1}{2} \, K)^2 \, \ell \right] \cdot \left[ -\lambda \, \delta_{\gamma\epsilon}\, (q-\frac{1}{2} \, K)^2 \, \ell \right], \quad \ell = \ln(1/m)
\end{equation}

\noindent using our initial approximation for the spin dependence of each loop. It should be noted that the color indices of the two sides of the loop are forced to be identical in the divergent limit of the loop.

In the absence of its $K$-dependence, Eq.~(\ref{Eq:13}) is just given by the product of the two loops' $q^2$-factors; and that $K$-dependence appears in the form of a sum over products of polynomial dependence on $K$ components, multiplying the transform $\widetilde{\mathcal{F}}$. Let us now take the inverse transform, writing
\begin{equation}
\widetilde{\mathcal{F}}(K) = \int{\mathrm{d}^{2}B \, e^{-i K \cdot B} \, [f \cdot \chi(B)]^{-1}},
\end{equation}

\noindent and noting that each $K$-component $K_{\alpha}$ can be expressed as a derivative with respect to $B_{\alpha}$  of the inverse transform, $K_{\alpha} \rightarrow i \frac{\partial}{\partial B_{\alpha}}$. An integration-by-parts transforms this derivative, and all such derivatives arising from the polynomial $K$-dependence of Eq.~(\ref{Eq:13}), into one or more derivatives operating upon $[ f \cdot \chi(B)]^{-1}$. But now the $\int{\mathrm{d}^{2} K}$ can be immediately evaluated, yielding $\delta^{(2)}(B)$, so that the result of all the $K$-dependence of Eq.~(\ref{Eq:13}) is a group of derivatives taken at $B=0$.

To evaluate this result, remember that we have not yet allowed the small space-time interval of the $(f\cdot \chi)^{-1}$ of this central GB to vanish, in conjunction with the loop UV divergences becoming infinite. Upon what portion of this small transverse volume do these $\frac{\delta}{\delta B_{\alpha}}$ operate? Those derivatives cannot have any bearing upon differences of this small volume and neighboring volumes, because each such small volume is completely independent of its neighbors. These derivatives refer to possible transverse variations within the small volume of interest, centered about the point $B=0$. But before taking its limit of zero volume, we are free to define how that limit is to be taken; and the only natural definition, and surely the simplest, is to imagine that volume as 'flat', without any curvature, so that each and every such derivative within that volume vanishes, after which the limit $B \rightarrow 0$ is taken. A better justification is simply that any curvature introduces a scale; and there is no relevant scale to adopt.

The contribution of this two-loop chain is then proportional to the product of two groups of $q$-factors, one from each loop, separated by the matrix quantity $[f\cdot \chi(0)]^{-1}|^{b'c'}_{\beta\gamma}$ which we now replace by the simplified expression
\begin{equation}\label{Eq:14}
(-\lambda q^2\delta_{\alpha\beta})(-\lambda q^2 \delta_{\gamma\epsilon}) \kappa^2.
\end{equation}

\noindent We leave it as an exercise for the interested reader to show that this result of the form of Eq.~(\ref{Eq:14}) will hold for every 'interior' GB of the chain. For example, following exactly the same procedures as for the two-loop amplitude above, the three-loop amplitude has four $(f\cdot \chi)^{-1}$ factors, and the central two may both be re-written as
\begin{equation}
\sum_{\beta, c'}{[f \cdot \chi(0)]^{-1}|^{b'c'}_{\alpha \beta} \cdot [ f \cdot \chi(0)]^{-1}|^{c'd'}_{\beta \gamma}}, \end{equation}

\noindent or as $[f\cdot \chi(0)]^{-2}|^{b'd'}_{\alpha \gamma}$.

In this way, the result for a chain with $n$ 'interior' GBs yields a term proportional to $\left. [ f\cdot \chi(0)]^{n} \right|^{b'd'}_{\alpha\gamma}$ which is inserted between the two 'exterior' GBs, $\left. [ f \cdot \chi(y_{\perp})]^{-1} \right|^{aa'}_{\mu\alpha}$ on the left and $\left. [f \cdot \chi(\bar{y}_{\perp})]^{-1} \right|^{d'b}_{\epsilon\nu}$ on the right, multiplied by the remaining $q$-dependence, and integrated over all transverse $q$. With $X = \lambda \, q^2 \, \kappa \, g \, \tilde{\varphi}(q)$, all together one has, upon summing over all interior loops which effectively form a geometric series, and including the amplitude with but one loop,
\begin{eqnarray}\label{Eq:15}
& & \left. [f \cdot \chi(y_{\perp})]^{-1}\right|^{aa'}_{\mu \alpha} \\ \nonumber
& & \quad  \cdot \, g X \tilde{\varphi} \, \left[ 1 + i X \, [f \cdot \chi(0)]^{-1} - X^2 \, [f \cdot \chi(0)]^{-2} - i X^3 \, [ f \cdot \chi(0)]^{-3} + \cdots \right]^{a'b'}_{\alpha\beta} \\ \nonumber
& & \quad \quad \cdot \left. [f \cdot \chi(\bar{y}_{\perp})]^{-1} \right|^{b'b}_{\beta\nu},
\end{eqnarray}

\noindent or, suppressing matrix indices,
\begin{equation}\label{Eq:16}
[ f \cdot \chi(y_{\perp}) ]^{-1} \cdot g  X \tilde{\varphi} \, \left[ \frac{1 + i X [f \cdot \chi(0)]^{-1}}{1 + X^2 \, [f \cdot \chi(0)]^{-2}} \right] \cdot [f\cdot \chi(\bar{y}_{\perp})]^{-1}.
\end{equation}

\noindent Eq.~(\ref{Eq:16}) can be replaced by
\begin{equation}\label{Eq:17}
g X \tilde{\varphi}\, [f\cdot \chi(y_{\perp})]^{-1} \cdot [f \cdot \chi(0)]^2 \,  \left[ \frac{ 1 + i X [f \cdot \chi(0)]^{-1}}{[f \cdot \chi(0)]^2 + X^2} \right] \cdot [f \cdot \chi(\bar{y}_{\perp})]^{-1}.
\end{equation}

Since the $\alpha$, $\beta$ indices of $\chi^{a}_{\alpha \beta}(0)$ are transverse, all components of $\chi(0)$ can be chosen as real; and since the $f^{abc}$ are also real, $[ f \cdot \chi(0)]^2$ is positive, and the denominator of Eq.~(\ref{Eq:17}) is never zero. The $\chi(0)$ contribution to the amplitude is then proportional to
\begin{equation}\label{Eq:18}
g^2 \, \int{\mathrm{d}^2 q \, e^{iq\cdot\vec{b}} \cdot [f \cdot \chi(y_{\perp})]^{-1} \cdot I(q^2,g^2) \cdot [f \cdot \chi(\bar{y}_{\perp})]^{-1}}
\end{equation}

\noindent or
\begin{equation}
g^2_R(q^2)=g^2 \ I(q^2,g^2) \ q^2 [\tilde{\varphi}(q)]^2 \ \lambda \kappa
\end{equation}

\noindent with
\begin{equation}\label{Eq:19}
I(q^2, g^2)=\mathcal{N} \, \int{\mathrm{d}^4 \chi(0) \, \det[f\cdot \chi(0)]^{-\frac{1}{2}} \,  e^{\frac{i}{4} \chi(0)^2} \, \frac{[f \cdot \chi(0)]^2}{[f \cdot \chi(0)]^2 + [\lambda \kappa g q^2 \tilde{\varphi}]^2} },
\end{equation}

\noindent since the integral $\int{\mathrm{d}^n \chi(0)}$ over an odd function of $[f \cdot \chi(0)]^{-1}$ vanishes.

While the integral of Eq.~(\ref{Eq:19}) may turn out to be complex, there is nothing really improper about a complex quantity multiplying any matrix element. To put this into a conventional form, where $g^2_R(q^2)$ is expected to be real, it may be possible to choose the product $\lambda \kappa$ so that $g_R$ can be made real; but the reality of such a $g_R(q^2)$ is an intuitive nicety, rather than a QM-requirement. In general, $g^2_R$ is a matrix quantity, and the same remarks applies.

One can see that there is no divergence in the integral of Eq.~(\ref{Eq:18}) for any value of $q^2$. The Fourier transform of this integral corresponds, in momentum space, to $q^2$ dependence -- following from this chain-graph form of renormalization -- of an effective, or renormalized, charge dependence of the complete set of radiative corrections obtained by summation over all contributing graphs. This factor will re-appear in every process describing interacting quarks; and as such, it can be considered as the effective, or renormalized color-charge dependence of this Model renormalization, as the 'renormalized' charge which appears in the scattering of a pair of quarks and/or anti-quarks, at $q$-values somewhat different from those obtained from simple one-GB exchange. Integrals over this quantity are then finite by virtue of the exponential cut-off appearing in $\tilde{\varphi}(q)$, which is (slightly) less strong than Gaussian, reflecting the basic structure of confinement in this Model of Realistic QCD. This chain-graph Bundle structure will be repeated in all of the correlation functions, with coordinates defined in terms of a basic CM frame; and while the integrations over coordinate components may become somewhat complicated, and require numerical integration, they are all finite.

Methods of Random Matrix theory~\cite{4,5}, requiring a certain measure of numerical computation, can be used to evaluate multiple-chain contributions to high-energy hadronic reactions, in particular elastic pp scattering. Our intention in the next few paragraphs is to demonstrate something much simpler -- the origin and appearance of the familiar "diffraction dip" in the $(\mbox{momentum-transfer})^2$ region of $\frac{m_p^2}{2}$ -- by adopting two intuitive, qualitative approximations for the exchange of a single GB-loop chain between a pair of scattering quarks, each bound into a different proton, with the details of that binding suppressed.

The first approximation is to represent the amplitude of a single chain by its first two terms, as pictured in Fig.~\ref{Fig:7}. The second approximation is to evaluate $\int{\mathrm{d}^{n} \chi(0)}$ by treating $\chi^a$ as a vector in color space, with magnitude $R=\sqrt{\sum_a(\chi^a)^2}$, greatly simplified by suppressing all of the normalized integrations over such angles, and retaining only the normalized integration over $R$.

With an arbitrary normalization of the absolute value of that $|\mbox{amplitude}|^2$, its value as a function of $q^2$ is represented in Fig.~\ref{Fig:8}. The dashed curve of Fig.~\ref{Fig:8}, largest at small $q^2$, is the result of using only a single GB exchange while the curve defined by dots represents the $|\mbox{amplitude}|^2$ of the single-loop exchange. The $|\mbox{amplitude}|^2$ of the sum of both the GB and the one-loop exchange is the solid curve of Fig.~\ref{Fig:8}, and easily displays the expected diffraction dip. When more loops are added to the total amplitude, the fall-off at larger $q^2$ should be reduced, while the dip should be moved very slightly to the right of ${m_p^2}/{2}$. This approximate evaluation of what appears to be the largest contributions to a process such as elastic pp scattering suggests that a detailed fit, fixing the as yet open parameters $g$, $\lambda \kappa$, $\delta_{q}(E)$ will at the very least be able to reproduce the essential features of the data. This project is now under detailed study.

\section{\label{SEC6}PERTURBATIVE AND NON-PERTURBATIVE APPROXIMATIONS}
%\subsection{}
%\subsubsection{}

Whether the chain-graph Model of QCD presented in the Previous Sections meets the essential criteria of experimental observation remains to be seen. It does permit a description of interactions between its fundamental quarks, from binding to scattering, and although the hadrons we measure are themselves bound states of quarks, the fall-off of measured hadronic scattering amplitudes with increasing momentum transfer has its counterpart in the 'renormalized' coupling constant of the chain-graph Model. If that fall-off turns out to be incorrect, then the Model must be discarded; but it does, at the very least, raise interesting questions about the structure, and comparison, of perturbative and non-perturbative theories.

Consider first QED, and in particular leptonic QED, in which photons are coupled to leptons which, not merely by definition but by experiment, seem to be fundamental, in the sense of having no sub-structure. The sequential calculations of their radiative corrections across the last half-century have always led to logarithmic UV divergences, whose sequential renormalization approximations -- that is, expressing all results in terms of the measurable, or renormalized charges and leptons masses to the same order of approximation -- have been shown to agree with experiment to a remarkable degree of accuracy. Nevertheless, as many authors have attempted to understand Ref.~\cite{10}, are there really divergences in QED, or is the appearance of such terms tied to the method of approximation?

That question has recently been partially answered in Ref.~\cite{friedgabelliniannals2012} by the application of a method of functional summation -- not just over a handful of Feynman graphs, but over infinite numbers of interactions, in each member of an infinite class, each containing an infinite number of Feynman graphs -- which strongly suggests that charge renormalization in QED is indeed finite. That analysis has not yet been extended to leptonic mass renormalization, nor to wave-function and vertex renormalization; but since the latter two quantities are equal and gauge-dependent, and always cancel in any physical measurement, one can accept their presence as an \textit{artefact} of calculation. Therefore, if gauge-independent charge renormalization is finite, one can accept QED not as an approximate theory, containing a still-hidden sub-structure, but rather as a True Theory of Nature.

Now consider QCD, which since its inception has always been defined in terms of a Lorentz-covariant Lagrangian, similar to but more complicated than those of Abelian theories. From the equations of motion of those theories, it is possible to define single-particle, asymptotic field operators whose quanta are described in terms of coordinates which can, in principle, be specified exactly, just as in QED. But we have known for several decades that quarks and antiquarks are always asymptotically bound to each other, and therefore that their transverse momenta and/or position coordinates can only, in principle, be described with quantum-mechanical precision, rather than specified exactly.

When the functional techniques referred to in the above Sections are applied to QCD, using a special rearrangement which guarantees manifest gauge invariance in Ref.~\cite{FriedGabellini2010}, the result of such a mismatch is the appearance of absurdities in non-perturbative amplitudes for all processes, divergences multiplying otherwise reasonable and finite factors. Once this most-inappropriate mismatch of the QM description is removed, as has been done phenomenologically in Ref.~\cite{2}, all such absurdities vanish, and one can see the essential difference of non-perturbative summations of a non-Abelian theory containing confinement, as compared to summations over one which violates that basic quantum-mechanical principle.

There is another difficulty with perturbative approximations in QCD, but one which could not have been known until summations over infinite numbers of gluons were functionally obtained. This is tied to the fact that the coupling constant $g$ appears in two different places in the Lagrangian, once as the coupling of quarks to gluons, and again in that part of the interaction coupling of gluons to each other. For definiteness, we shall refer to the first coupling as $g_1$, and to the second as $g_2$; they are, of course, to be set equal to each other, but it will be instructive to keep the distinction for a few more lines.

The summation of all gluon exchanges between any two quark lines leads to amplitudes depending upon factors of $\frac{g_1^2}{g_2}$; and at this stage one can see the difficulties which arise when attempting perturbative expansions in $g$, that is, at this stage in $g_1$ and $g_2$, for one cannot expand a function of $\frac{1}{g_2}$ about $g_2=0$. Such an expansion of both $g_1$ and $g_2$ corresponds to treating QCD as a mixture of QED and Yang-Mills, and all points in-between; and leads to irrelevant divergences and confusion. This could not have been foreseen until the summations over all gluon exchanges were performed, but it illustrates the difficulties of a purely perturbative approach to QCD.

Finally, it may be useful to consider perturbative approximations of radiative corrections to particles which are themselves bound states of more fundamental objects, such as nucleons are of quarks, in mock-Abelian theories, such as QED applied directly to protons. The simplest self-energy Feynman graph of a proton emitting and re-absorbing a photon, illustrates the point: Why is there a UV divergence associated with this graph? The answer is simply because it has been tacitly assumed that the proton still exists, after emitting and then re-absorbing a virtual photon of sufficiently high energy, which produced the UV divergence.

Alternatively, consider the absorptive part of that amplitude, corresponding to the absorption of a photon by the struck proton, and the emission of the final photon by the proton. If the initial photon energy is far less than the binding energy of that three-quark proton state, then the Feynman graph is perfectly relevant, and the contribution of the corresponding dispersion relation to the self-energy graph is finite. But when that photon's energy is far greater than the 3-quark binding energy, it will split the proton apart into its three fundamental quarks, which, according to the chain-Model of this paper -- will yield a perfectly finite amplitude, as will that of the re-combinations of each of those quarks into whichever asymptotic states they may form themselves. Again, it is the tacit assumption that the proton is itself a fundamental particle which leads to the UV divergence; and the removal of that and all other such UV divergences is simply to insert a cut-off statement, into the basic Lagrangian, carrying the information that the Lagrangian is only true when interacting photon energies are less than the proton's bound-state energy.

\section{\label{SEC7}SUMMARY}
%\subsection{}
%\subsubsection{}

These next paragraphs are not intended to be a restatement of previously mentioned items above, but rather a final insertion of a few points previously not emphasized.

It may have been overlooked, but the final form of $\mathbf{L}[A]$ -- after its A-dependence has been translated so as to incorporate the $u'$ and $\bar{u}'$ variables of the two quark lines -- is in an exponential, along with all the $(f\cdot \chi)^{-1}$ of all GBs. (The Halpern integrals over the three different $(f\cdot \chi)^{-1}$ are non exponentiated.) The expansion of that exponential in its powers of $u'$ and $\bar{u}'$ corresponds to the interacting quarks exchanging more than one set of GB chain graphs -- but chains whose loop substructures can never interact with each other, since such loop interactions via new GBs would vanish under this Model renormalization. Whether more than one complete GB chain should be considered would depend upon how much time is allowed for any such reaction, how fast the quarks are moving, etc. It would allow the different chains to interfere with each other, in a QM way but as \underline{entities}, and not allow their loop-sub-structures to interact with each other.

The complete loop-exchange functional structure here has the form
\begin{eqnarray}\label{Eq:20}
&& \int{\mathrm{d}^n \, \chi(y_{\perp}) \det[f\cdot \chi(y_{\perp})]^{{-\frac{1}{2}}} \, e^{{\frac{i}{4}}\chi^{2}(y_{\perp})}} \\ \nonumber
&& \cdot \int{\mathrm{d}^n \, \chi(\widebar{y}_{\perp}) \det[f\cdot \chi(\widebar{y}_{\perp})]^{{-\frac{1}{2}}} \, e^{{\frac{i}{4}} \chi^{2}(\widebar{y}_{\perp})} } \\ \nonumber
&& \cdot \int{\mathrm{d}^n \, \chi(0) \det[f\cdot \chi(0)]^{{-\frac{1}{2}}} \, e^{{\frac{i}{4}} \chi^2(0)}}
\\ \nonumber
&& \cdot \exp{\left[ \int_{0}^{s}{\mathrm{d}s_1 \, u'_{\mu}(s_1) \Omega_a(s_1) \left.(f \cdot \chi(y_{\perp}))^{-1} \right|^{aa'}_{\alpha\mu} } \right] } \, \frac{\delta}  {\delta A^{a'}_{\alpha}(y-u(s_1))}
\\ \nonumber
&& \cdot \exp{ \left[ \int_{0}^{\bar{s}}{\mathrm{d}\bar{s}_1 \, \bar{u}'_{\nu}(\bar{s}_1)  \widebar{\Omega}_b(\bar{s_1}) \left. (f\cdot \chi(\bar{y}_{\perp}))^{-1} \right|^{b'b}_{\beta\nu} } \right] } \, \frac{\delta}{\delta A^{b}_{\beta}(\bar{y}-\bar{u}(\bar{s}_1))} \\ \nonumber
&& \cdot \exp{\left\{ \mathbf{L}[A] + \mathbf{L}[A] \, (e^{\overleftrightarrow{\mathcal{D}}} - 1) \, \mathbf{L}[A]+ \cdots \right\}}.
\end{eqnarray}

\noindent Each of the $\mathbf{L}[A]$ entering into Eq.~(\ref{Eq:20}) is then to have an $A$-dependence, which itself enters in the form of an exponential, shifted by the translation operators of the $u$ and $\bar{u}$ quantities, such that after the individual loop-functional integrations are performed, the result will be the exponential factor
\begin{equation}
\exp{\left[ \int{\mathrm{d}^2 q \, e^{i\vec{q}\cdot\vec{b}} \, g \, X \, \tilde{\varphi} \, \left( \frac{1 + iX \, [f \cdot \chi(0)]^{-1}}{1 + X^2 \, [f\cdot \chi(0)]^{-2}} \right) } \right]},
\end{equation}

\noindent so that the entire multiple GB chain contribution to the scattering -- in which none of the sub-elements of any chain can interact with those of another chain -- takes the form
\begin{eqnarray}\label{Eq:21}
&& \exp{ \left\{ \int_{0}^{s}{\mathrm{d}s_1 \, u'_{\mu}(s_1) \, \Omega_a(s_1) \left. [f \cdot \chi(y_{\perp})]^{-1} \right|^{aa'}_{\mu \alpha}} \right. }  \\ \nonumber
&& \quad \cdot \int{\mathrm{d}^{2}q \, e^{i\vec{q}\ \vec{b}} g\ X\ \widetilde{\varphi} \left.\left(\frac{1+iX(f\cdot \chi(0))^{-1}}{1+X^2(f\cdot \chi(0))^{-2}}\right)\right|^{a'b'}_{\alpha\beta}} \\ \nonumber
&& \quad \cdot \left. \int_0^{\bar{s}}{\mathrm{d}\bar{s}_1 \, \bar{u}'_{\mu}(\bar{s}_1) \, \widebar{\Omega}_b(\bar{s}_1) \left. [f \cdot \chi(\bar{y}_{\perp})]^{-1}\right|_{\beta\nu}^{b'b}} \right\}
\end{eqnarray}

\noindent Retaining only the linear terms in $u'$ and $\bar{u}'$ of Eq.~(\ref{Eq:21}) corresponds to the exchange of a single GB chain, as in Section~\ref{SEC4}.

The simplest interpretation, consistent with fast-moving quarks described by an eikonal representation, is obtained by expanding the $u'$ and $\bar{u}'$, retaining only linear $u'$ and $\bar{u}'$ dependence -- and therefore their own $(f\cdot \chi)^{-1}$ variables, and thereby bringing down the entire $(f\cdot \chi)^{-1}$ dependence, as written in this paper -- so that only one GB chain is exchanged. That analysis is sufficient to produce $(g_R)^2$ as a function of $q^2$, which quantity may be complex and matrix-valued.

Two things are important, and quite attractive, about this Model QCD renormalization:
\begin{enumerate}
\item The fact that everything comes out finite, as each loop's UV divergence is absorbed by the vanishing $(\delta)^2$ of the Halpern FI; and because the final integral over $d[\chi(0)] $ should be perfectly finite, even with the $(f\cdot \chi(0))^{-1}$ terms all up (as in the first paragraph above) in the exponential. It can be easily estimated by suppressing all "angular" color and tranverse coordinate dependence, and simply integrating over "the magnitude" of $R$ of $(f\cdot \chi(0))^{-1}$; and it produces, as expected, a strong dependence on $g^2 X\ \widetilde{\varphi}^2$, especially for large $q^2$, so that one finds a strong fall-off with increasing $q^2$ for the effective, or 'renormalized' charge, just as expected and needed.  This can easily be seen by using Random Matrix methods, as in Ref.~\cite{4,5}.

\item Just as one may now have confidence in QED as being a "fundamental and true Theory of Nature", because its charge renormalization is almost surely finite~\cite{friedgabelliniannals2012}, and there is no need to hunt for any 'underlying' Theory which could magically produce that effect, so may QCD be tentatively called a "fundamental and true Theory of Nature"~\cite{wilczek}, at least in this simplest renormalization Model. Whether or not the $q^2$ dependence derived for quark-quark interactions, when incorporated into hadron scattering and production processes, turns out to be that required by experiment, is the crucial point. If so, then this model will have the right to be assumed correct and proper. If not, the experience gained with this functional approach will suggest that somewhat more complicated calculations must be done, difficult but certainly possible, before QCD can be placed in the same, high category as QED.
\end{enumerate}

\begin{acknowledgments}
This publication was made possible through the partial support of a Grant from the Julian Schwinger Foundation. The opinions expressed in this publication are those of the authors and do not necessarily reflect the views of the Julian Schwinger Foundation. We especially wish to thank Mario Gattobigio, of Universit\'{e} de Nice Sophia-Antipolis, for his many kind and informative conversations relevant to the Nuclear Physics aspects of our work. It is also a pleasure to thank Mark Restollan, of the American University of Paris, for his kind assistance in arranging sites for our collaborative research when in Paris.
\end{acknowledgments}

\appendix

\section{\label{AppA}Fradkin's Representations for Green's Function and Closed-Fermion-Loop Functional}
%\section{\label{AppA}Fradkin's Representations for $\mathbf{G}_{\mathrm{c}}[A]$ and $\mathbf{L}[A]$}

The exact functional representations of these two functionals of $A(x)$ are perhaps the most useful tools in all of QFT, for they allow that $A$-dependence of these functionals to be extracted from inside ordered exponentials; and because they, themselves, are Gaussian in their dependence upon $A(x)$, they permit the functional operations of the Schwinger/Symanzik generating functional (Gaussian functional integration, or functional linkage operation) to be performed exactly.  This corresponds to an explicit sum over all Feynman graphs relevant to the process under consideration, with the results expressed in terms of functional integrals over the Fradkin variables; and in the present QCD case, because of EL, those non-perturbative results can be extracted and related to physical measurements.

The causal quark Green's function (which is essentially the most customary Feynman one) can be written as~\cite{8,9}
\begin{equation}\label{Eq:A1}
\mathbf{G}_{c}[A] = [ m  + i \gamma \cdot \Pi][m  + (\gamma \cdot \Pi)^{2}]^{-1} = [ m  + i \gamma \cdot \Pi] \cdot i \int_{0}^{\infty}{ds \, e^{-ism^{2}} \, e^{is(\gamma \cdot \Pi)^{2}}  },
\end{equation}

\noindent where $\Pi = i [\partial_{\mu} - i g A_{\mu}^{a} \tau^{a}]$ and $(\gamma \cdot \Pi)^{2} = \Pi^{2} + i g \sigma_{\mu \nu} \, \mathbf{F}_{\mu \nu}^{a} \tau^{a}$ with $\sigma_{\mu \nu} = \frac{1}{4} [\gamma_{\mu}, \gamma_{\nu}]$.  Following Fradkin's method \cite{8,9} and replacing $\Pi_{\mu}$ with $i \frac{\delta}{\delta v_{\mu}}$, one obtains
\begin{eqnarray}\label{Eq:A2}
& & \mathbf{G}_{\mathrm{c}}(x,y|A) \\ \nonumber &=& i \int_{0}^{\infty}{ds \ e^{-ism^{2}} \cdot e^{i \int_{0}^{s}{ds'
\frac{\delta^{2}}{\delta v_{\mu}^{2}(s')}}} \cdot \left[ m - \gamma_{\mu} \, \frac{\delta}{\delta v_{\mu}(s)} \right] } \, \delta( x -y + \int_{0}^{s}{ds' \ v(s')}) \\ \nonumber & &
\times \left.  \left(\exp{\left\{ -ig \int_{0}^{s}{ds' \left[v_{\mu}(s') \, A_{\mu}^{a}(y-\int_{0}^{s'}{v}) \tau^{a} + i \sigma_{\mu \nu} \, \mathbf{F}_{\mu \nu}^{a}(y-\int_{0}^{s'}{v}) \tau^{a} \right] }\right\}} \right)_{+}  \right|_{v_{\mu} \rightarrow 0}.
\end{eqnarray}

\noindent Then, one can insert a functional `resolution of unity' of form
\begin{equation}\label{Eq:A3}
1 = \int{\mathrm{d}[u] \, \delta[u(s') - \int_{0}^{s'}{ds'' \ v(s'')}]},
\end{equation}

\noindent and replace the delta-functional $\delta[u(s') - \int_{0}^{s'}{ds'' \ v(s'')}]$ with a functional integral over $\Omega$, and then the Green's function becomes~\cite{YMS2008}
\begin{eqnarray}\label{Eq:A4}
&& \mathbf{G}_{\mathrm{c}}(x,y|A) \\ \nonumber &=&  i \int_{0}^{\infty}{ds \ e^{-is m^{2}}} \, e^{- \frac{1}{2} \Tr{\ln{\left( 2h \right)}} } \, \int{d[u]} \, e^{ \frac{i}{4} \int_{0}^{s}{ds' \, [u'(s')]^{2} } } \, \delta^{(4)}(x - y + u(s)) \\ \nonumber & & \quad \times {\left[ m + i g \gamma_{\mu} A_{\mu}^{a}(y-u(s)) \tau^{a} \right]} \, \left( e^{ -ig \int_{0}^{s}{ds' \, u'_{\mu}(s') \, A_{\mu}^{a}(y-u(s')) \, \tau^{a}} + g \int_{0}^{s}{ds' \sigma_{\mu \nu} \, \mathbf{F}_{\mu \nu}^{a}(y-u(s')) \, \tau^{a}}} \right)_{+},
\end{eqnarray}

\noindent where $h(s_{1},s_{2})=\int_{0}^{s}{ds' \, \Theta(s_{1} - s') \Theta(s_{2} - s')}$.  To remove the $A$-dependence out of the linear (mass) term, one can replace $i g A_{\mu}^{a}(y-u(s)) \tau^{a}$ with $- \frac{\delta}{\delta u'_{\mu}(s)}$ operating on the ordered exponential so that
\begin{eqnarray}\label{Eq:A5}
&& \mathbf{G}_{\mathrm{c}}(x,y|A) \\ \nonumber &=&  i \int_{0}^{\infty}{ds \ e^{-is m^{2}}} \, e^{- \frac{1}{2} \Tr{\ln{\left( 2h \right)}} } \, \int{d[u]} \, e^{ \frac{i}{4} \int_{0}^{s}{ds' \, [u'(s')]^{2} } } \, \delta^{(4)}(x - y + u(s)) \\ \nonumber & & \quad \times {\left[ m - \gamma_{\mu} \frac{\delta}{\delta u'_{\mu}(s)} \right]} \, \left( e^{ -ig \int_{0}^{s}{ds' \, u'_{\mu}(s') \, A_{\mu}^{a}(y-u(s')) \, \tau^{a}} + g \int_{0}^{s}{ds' \sigma_{\mu \nu} \, \mathbf{F}_{\mu \nu}^{a}(y-u(s')) \, \tau^{a}}} \right)_{+}.
\end{eqnarray}

\noindent To extract the $A$-dependence out of the ordered exponential, one may use the following identities,
\begin{eqnarray}
1 &=& \int{d[\alpha] \, \delta{\left[ \alpha^{a}(s') + g u'_{\mu}(s') \, A^{a}_{\mu}(y-u(s'))\right]}},  \\ \nonumber 1 &=& \int{d[\mathbf{\Xi}] \, \delta{\left[ \mathbf{\Xi}^{a}_{\mu \nu}(s') - g \mathbf{F}_{\mu \nu}^{a}(y-u(s'))\right]} },
\end{eqnarray}

\noindent and the ordered exponential becomes
\begin{eqnarray}
& & \left( e^{ -ig \int_{0}^{s}{ds' \, u'_{\mu}(s') \, A_{\mu}^{a}(y-u(s')) \, \tau^{a}} + g \int_{0}^{s}{ds' \sigma_{\mu \nu} \, \mathbf{F}_{\mu \nu}^{a}(y-u(s')) \, \tau^{a}}} \right)_{+} \\ \nonumber &=& \mathcal{N}_{\Omega} \, \mathcal{N}_{\Phi} \, \int{d[\alpha] \, \int{d[\mathbf{\Xi}] \, \int{d[\Omega] \, \int{d[\mathbf{\Phi}] \, \left( e^{ i \int_{0}^{s}{ds' \, \left[ \alpha^{a}(s') - i \sigma_{\mu \nu} \, \mathbf{\Xi}_{\mu \nu}^{a}(s') \right] \, \tau^{a}}} \right)_{+} }}}} \\ \nonumber & & \quad \times e^{-i \int{ds' \, \Omega^{a}(s') \, \alpha^{a}(s')}  - i \int{ds' \, \mathbf{\Phi}^{a}_{\mu \nu}(s') \,  \mathbf{\Xi}^{a}_{\mu \nu}(s') } } \\ \nonumber & & \quad \times e^{- i g \int{ds' \, u'_{\mu}(s') \, \Omega^{a}(s') \, A^{a}_{\mu}(y-u(s')) } + i g \int{ds' \, \mathbf{\Phi}^{a}_{\mu \nu}(s') \, \mathbf{F}_{\mu \nu}^{a}(y-u(s'))}  },
\end{eqnarray}

\noindent where $\mathcal{N}_{\Omega}$ and $\mathcal{N}_{\Phi}$ are constants that normalize the functional representations of the delta-functionals.  All $A$-dependence is removed from the ordered exponential and the resulting form of the Green's function is exact (it entails no approximation).  Alternatively, extracting the $A$-dependence out of the ordered exponential can also be achieved by using the functional translation operator, and one writes
\begin{eqnarray}
&& \left( e^{ + g \int_{0}^{s}{ds' \, \left[ \sigma_{\mu \nu} \, \mathbf{F}_{\mu \nu}^{a}(y-u(s')) \tau^{a} \right]} }\right)_{+} \\ \nonumber
&& = \left. e^{g \int_{0}^{s}{ds' \, \mathbf{F}_{\mu \nu}^{a}(y-u(s')) \, \frac{\delta}{\delta \mathbf{\Xi}_{\mu\nu}^{a}(s')}}} \cdot \left( e^{\int_{0}^{s}{ds' \, \left[ \sigma_{\mu \nu} \, \mathbf{\Xi}_{\mu \nu}^{a}(s') \tau^{a} \right]} }\right)_{+} \right|_{\mathbf{\Xi} \rightarrow 0}.
\end{eqnarray}

For the closed-fermion-loop functional $\mathbf{L}[A]$, one can write~\cite{9}
\begin{equation}\label{Eq:ClosedFermionLoopFunctional02}
\mathbf{L}[A] = - \frac{1}{2} \, \int_{0}^{\infty}{\frac{ds}{s} \, e^{-ism^{2}} \, \left\{\Tr{\left[ e^{-is(\gamma \cdot \Pi)^{2}} \right]} - \left\{ g=0 \right\} \right\}},
\end{equation}

\noindent where the trace $\Tr{}$ sums over all degrees of freedom, space-time coordinates, spin and color.  The Fradkin representation proceeds along the same steps as in the case of $\mathbf{G}_{\mathrm{c}}[A] $, and the closed-fermion-loop functional reads
\begin{eqnarray}\label{Eq:LFradkin01}
\mathbf{L}[A] &=&  - \frac{1}{2} \int_{0}^{\infty}{\frac{ds}{s} \, e^{-is m^{2}}} \, e^{- \frac{1}{2} \Tr{\ln{(2h)}} } \\ \nonumber && \quad \times \int{d[v]} \, \delta^{(4)}(v(s)) \, e^{ \frac{i}{4} \int_{0}^{s}{ds' \, [v'(s')]^{2} } } \\ \nonumber & & \quad \times \int{d^{4}x \, \tr{\left( e^{ -ig \int_{0}^{s}{ds' \, v'_{\mu}(s') \, A_{\mu}^{a}(x-v(s')) \, \tau^{a}} + g \int_{0}^{s}{ds' \sigma_{\mu \nu} \, \mathbf{F}_{\mu \nu}^{a}(x-v(s')) \, \tau^{a}}} \right)_{+}} } \\ \nonumber & & - \left\{ g = 0 \right\},
\end{eqnarray}

\noindent where the trace $\tr{}$ sums over color and spinor indices.  Also, Fradkin's variables have been denoted by $v(s')$, instead of $u(s')$, in order to distinguish them from those appearing in the Green's function $\mathbf{G}_{\mathrm{c}}[A] $. One finds
\begin{eqnarray}\label{Eq:LFradkin02}
\mathbf{L}[A] &=&  - \frac{1}{2} \int_{0}^{\infty}{\frac{ds}{s} \, e^{-is m^{2}}} \, e^{- \frac{1}{2} \Tr{\ln{(2h)}} } \\ \nonumber && \quad \times \mathcal{N}_{\Omega} \, \mathcal{N}_{\Phi} \int{d^{4}x \, \int{\mathrm{d}[\alpha] \, \int{\mathrm{d}[\Omega] \, \int{\mathrm{d}[\mathbf{\Xi}] \, \int{\mathrm{d}[\mathbf{\Phi}] \, }}}}}  \\ \nonumber & & \quad \times \int{d[v] \, \delta^{(4)}(v(s)) \, e^{ \frac{i}{4} \int_{0}^{s}{ds' \, [v'(s')]^{2} } }  } \\ \nonumber & & \quad \times \ e^{-i \int{ds' \, \Omega^{a}(s') \, \alpha^{a}(s')} - i \int{ds' \, \mathbf{\Phi}^{a}_{\mu \nu}(s') \,  \mathbf{\Xi}^{a}_{\mu \nu}(s') } }  \cdot  \tr{\left( e^{ i \int_{0}^{s}{ds' \, \left[ \alpha^{a}(s') - i \sigma_{\mu \nu} \, \mathbf{\Xi}_{\mu \nu}^{a}(s') \right] \, \tau^{a}}} \right)_{+}} \\ \nonumber & & \quad \times  e^{- i g \int_{0}^{s}{ds' \, v'_{\mu}(s') \, \Omega^{a}(s') \, A^{a}_{\mu}(x - v(s'))} - 2 i g \int{d^{4}z \, \left(\partial_{\nu}  \mathbf{\Phi}^{a}_{\nu \mu}(z) \right) \, A^{a}_{\mu}(z) }} \\ \nonumber & & \quad \times e^{ + i g^{2} \int{ds' \, f^{abc} \mathbf{\Phi}^{a}_{\mu \nu}(s') \, A^{b}_{\mu}(x- v(s')) \, A^{c}_{\nu}(x- v(s')) }}  \\ \nonumber & & - \left\{ g = 0 \right\},
\end{eqnarray}

\noindent where the same properties as those of $\mathbf{G}_{\mathrm{c}}[A] $ can be read off readily.

\section{\label{AppB} Vanishing 'Bundle Self-Energy' Diagrams of Fig.~1 and 2.}

For simplicity and clarity, we first consider the non-spin dependence of the Fradkin representation of $\mathbf{G}_{\mathrm{c}}[A]$, and then discuss the spin terms separately. Because of the \emph{Effective Locality} (EL), the 4-dimensional delta-function multiplying the $(f \cdot \chi)^{-1}$ factor of the GB of Fig.~\ref{Fig:1} is given by $\delta^{(4)}(u(s_1)-u(s_2))$. This suggests but does not necessarily require that $s_1 = s_2$; but that condition is obtained by considering the time-like and longitudinal integrals separately, $\delta(u_0(s_1)-u_0(s_2))$ and $\delta(u_L(s_1)-u_L(s_2))$. Suppose now that there are a set of points $s_\ell$ for which the argument of the time-like $\delta_{(0)}$ vanishes, and a set of points $s_m$ for which the argument of the longitudinal $\delta_{(L)}$-function vanishes,
\begin{eqnarray}
\delta_{(0)} &=& \sum_{\ell} \frac{\delta(s_1-s_{\ell})}{|u'_0(s_{\ell})|} \bigg|_{u_0(s_{\ell})=u_0(s_2)},
\\ \nonumber
\delta_{(L)} &=& \sum_{m} \frac{\delta(s_s-s_{m})}{|u'_L(s_{m})|} \bigg|_{u_L(s_{1})=u_L(s_m)}.
\end{eqnarray}

\noindent Their product is then given by
\begin{eqnarray}
\sum_{\ell,m}  \frac{\delta(s_1-s_{\ell})\delta(s_1-s_m)}{|u'_0(s_{\ell})\ u'_L(s_m)|}
\bigg|_{ \substack{u_0(s_{\ell})=u_0(s_m)\\ u_L(s_{\ell})=u_L(s_m)}  },
\end{eqnarray}

\noindent and it is the subsidiary conditions which are most relevant. Since $u_0$ and $u_L$ are continuous but completely independent functions, the probability of finding sets of points $s_{\ell}$ and $s_m$ at which $u_0$ takes on the same value, and at which $u_L$ simultaneously has the same value, would appear to be less than $\epsilon$, and $\epsilon \rightarrow 0$. However, there are two $s$-values for which this is possible, where initial conditions specify that $u_{\mu}(0)=0$, and that $u_{\mu}(s)=-z_{\mu}$. Therefore, only $s_1 = s_2 = 0$, or else $s_1 = s_2 = s$. Then, for either case, $s_1 = s_2$, and the coefficients $u'_{\mu}(s_1)$ and $u'_{\nu}(s_2)$ are symmetric in $\mu$ and $\nu$, and are multiplying $(f\cdot \chi)^{-1}|_{\mu\nu}$ which is antisymmetric in those indices; and the result is zero.

The spin dependence for this particular process will also vanish, but for two different reasons. Those terms coming from the linear $A$-dependence of the $\mathbf{G}_{\mathrm{c}}[A]$ representation will have gradient terms differentiating the $y$-dependence of the $\delta$-functions representing \emph{EL}, but that $y$-dependence trivially cancels for this 'self-energy' process, and hence those terms give a zero result. The antisymmetric spin dependence coming from quadratic $A$-terms finds itself multiplying a different set of $u'_{\mu}(s_1)$ and $u'_{\nu}(s_2)$ coefficients; and then the analysis of the previous paragraph again rules out any non-zero contribution.

The vanishing of the Bundle Diagram of Fig.~\ref{Fig:2} may be inferred from that of Fig.~\ref{Fig:1}, by imagining the two ends of the quark line of Fig.~\ref{Fig:1} to be wrapped around and form a closed loop; and then, without performing the loop integrations, the result is zero. Or, one may follow the argument used in the text following Eq.~(\ref{Eq:6}) for chain-graph loops but applied to this single loop containing an internal GB; and again the result is zero.


\end{document}